\documentclass[reprint,twocolumn,superscriptaddress,showpacs,showkeys,amsmath,amssymb,floatfix,aps,prb]{revtex4-1}
\usepackage[english]{babel}
\usepackage[utf8]{inputenc}
\usepackage{graphicx}
\usepackage{fancyhdr}
\usepackage[active]{srcltx}
\usepackage{setspace}
\usepackage{float}
\usepackage{color}
\usepackage{multirow}
\usepackage{gensymb}

\usepackage[hidelinks]{hyperref}
\usepackage{xcolor}
\hypersetup{
    colorlinks,
    linkcolor={red},
    citecolor={blue},
    urlcolor={magenta}
}

\newcommand{\MS}{MoSe$_\textrm{2}$}
\newcommand{\WS}{WSe$_\textrm{2}$}
\newcommand{\Rh}{R$^{\textrm{h}}_{\textrm{h}}$}
\newcommand{\RM}{R$^{\textrm{M}}_{\textrm{h}}$}
\newcommand{\RX}{R$^{\textrm{X}}_{\textrm{h}}$}
\newcommand{\Hh}{H$^{\textrm{h}}_{\textrm{h}}$}
\newcommand{\HM}{H$^{\textrm{M}}_{\textrm{h}}$}
\newcommand{\HX}{H$^{\textrm{X}}_{\textrm{h}}$}

\begin{document}

\title{Signatures of electric field and layer separation effects on the spin-valley physics of \MS/\WS~heterobilayers: from energy bands to dipolar excitons}

\author{Paulo E. Faria~Junior}
\email{fariajunior.pe@gmail.com; \\ paulo-eduardo.faria-junior@ur.de}
\affiliation{Institute for Theoretical Physics, University of Regensburg, 93040 Regensburg, Germany}

\author{Jaroslav Fabian}
\affiliation{Institute for Theoretical Physics, University of Regensburg, 93040 Regensburg, Germany}

\date{\today}

%================================================================================

\begin{abstract}

Multilayered van der Waals heterostructures based on transition metal dichalcogenides are suitable platforms to study interlayer (dipolar) excitons, in which electrons and holes are localized in different layers. Interestingly, these excitonic complexes exhibit pronounced valley Zeeman signatures. However, it remains largely unexplored how their spin-valley physics can be further altered due to external parameters, such as electric field and interlayer separation. Here, we perform a systematic analysis of the spin-valley physics in \MS/\WS~heterobilayers under the influence of an external electric field and changes of the interlayer separation. Particularly, we analyze the spin ($S_z$) and orbital ($L_z$) degrees of freedom, as well as the symmetry properties of the relevant band edges (at K, Q, and $\Gamma$ points) of high-symmetry stackings at 0\degree (R-type) and 60\degree (H-type) angles, the important building blocks present in moiré or atomically reconstructed structures. We reveal distinct hybridization signatures on the spin and the orbital degrees of freedom of low-energy bands due to the wave function mixing between the layers, which are stacking-dependent and can be further modified by electric field and interlayer distance variation. We found that H-type stackings favor large changes in the g-factors as a function of the electric field, e. g. from $-5$ to $3$ in the valence bands of the \Hh~stacking, because of the opposite orientation of $S_z$ and $L_z$ of the individual monolayers. For the low-energy dipolar excitons (direct and indirect in $k$-space), we quantify the electric dipole moments and polarizabilities, reflecting the layer delocalization of the constituent bands. Furthermore, our results show that direct dipolar excitons carry a robust valley Zeeman effect nearly independent of the electric field but tunable by the interlayer distance, which can be experimentally accessible via applied external pressure. For the momentum-indirect dipolar excitons, our symmetry analysis indicates that phonon-mediated optical processes can easily take place. Particularly for the indirect excitons with conduction bands at the Q point for H-type stackings, we found marked variations of the valley Zeeman ($\sim 4$) as a function of the electric field that notably stand out from the other dipolar exciton species. Our analysis suggest that stronger signatures of the coupled spin-valley physics are favored in H-type stackings, which can be experimentally investigated in samples with twist angle close to 60\degree. In summary, our study provides fundamental microscopic insights into the spin-valley physics of van der Waals heterostructures that are relevant to understand the valley Zeeman splitting of dipolar excitonic complexes and also \textit{intralayer} excitons. 

\end{abstract}

\keywords{Valley Zeeman, spin-valley, van der Waals heterostructure, TMDC}

\maketitle

%================================================================================

\section{Introduction}

Transition metal dichalcogenides (TMDCs) monolayers host a plethora of fascinating physical properties. For instance, these materials are direct band gap semiconductors with strong spin-orbit coupling (SOC)\cite{Mak2010PRL, Splendiani2010NL, Kuc2011PRB, Kumar2012EPJB, Xiao2012PRL} and display robust signatures of excitonic and valley contrasting physics\cite{Mak2012NatNano, Mak2014Science, Qiu2013PRL, Xu2014NatPhys, Chernikov2014PRL, Stier2016NatComm, Schaibley2016NatRevMat, Raja2017NatComm, Wang2018RMP}. Furthermore, the van der Waals nature of TMDC materials facilitates the vertical stacking and integration of different layers, the key ingredient driving the burgeoning research field of van der Waals heterostructures\cite{Geim2013, Song2018, Sierra2021NatNano, Ciarrocchi2022NatRevMat}. These novel heterostructures offer a fantastic playground to investigate emergent physical phenomena and effects of proximity interaction, certainly not limited to TMDCs\cite{Tran2019Nature, Seyler2019Nature, Leisgang2020NatNano, Lin2021NatComm_bi} but also encompassing other attractive materials, such as graphene\cite{Gmitra2015PRB, Gmitra2016PRB, Avsar2017ACSNano, Luo2017NL, Raja2017NatComm, FariaJunior2023}, boron nitride\cite{Cadiz2017PRX, Goryca2019NatComm, Yankowitz2019NatRevPhys}, ferromagnetic CrX$_3$ (X=I,Br)\cite{Zollner2019PRBcri3, Zollner2023PRB, Zhong2017SciAdv, Ciorciaro2020PRL, Choi2022NatMat} and many others.

Particularly for the TMDC-based van der Waals heterostructures, one relevant system with great interest is the \MS/\WS~heterobilayer. In these systems, the type-II band alignment favors the appearance of interlayer excitons, also known as dipolar excitons, with electrons and holes localized in different layers\cite{Chaves2018PRB, Gillen2018PRB, Torun2018PRB}. Interestingly, dipolar excitons exhibit long recombination times\cite{Rivera2015NatComm, Miller2017NL, Nagler2017TDM}, robust out-of-plane electric dipole moments\cite{Ciarrocchi2019NatPhot, Baek2020SciAdv, Shanks2021NL, Barre2022Science, Shanks2022NL}, and giant valley Zeeman signatures that are dependent on the twist angle and atomic registry\cite{Nagler2017NatComm, Seyler2019Nature, Baek2020SciAdv, Mahdikhanysarvejahany2021npj, Holler2022PRB, Smirnov2022TDM, Li2022arXiv}. These are attractive features that could be exploited in exciton-based opto-spintronics devices\cite{Sierra2021NatNano, Ciarrocchi2022NatRevMat}, such as dipolar excitons quantum emitters\cite{Baek2020SciAdv} and diode transistors\cite{Shanks2022NL}. Furthermore, these \MS/\WS~are suitable platforms to study the optical signatures of moiré superlattices\cite{Tran2019Nature, Seyler2019Nature, Choi2021PRL, Forg2021NatComm} and atomic reconstruction\cite{Enaldiev2020PRL, Holler2020APL, Parzefall2021TDM, Zhao2022}. From a fundamental perspective, there is still some debate on the energy bands that constitute these dipolar excitons. A few reports suggest momentum-indirect dipolar excitons originating from valence bands at the K point and conduction bands at the Q point\cite{Miller2017NL, Hanbicki2018ACSNano, Sigl2020PRR}, despite the strong evidence of momentum-direct dipolar excitons with conduction and valence bands located at the K point\cite{Nagler2017NatComm, Seyler2019Nature, Wang2020NL, Delhomme2020TDM, Mahdikhanysarvejahany2021npj, Holler2022PRB, Smirnov2022TDM, Li2022arXiv} with a distinct valley Zeeman signature of $\sim -16$. Interestingly, it has recently been shown that these different dipolar exciton species exhibit different electric dipole moments\cite{Barre2022Science}, thus providing a possible alternative to properly distinguish them under applied electric fields. In order to fully exploit the functionalities of dipolar excitons in van der Waals heterostructures, it is particularly important to understand their spin-valley physics and how it depends on external parameters, such as applied electric fields\cite{Ramasubramaniam2011PRB, Chaves2018PRB} and fluctuations of the interlayer distance\cite{Plankl2021NatPhot}.

In this manuscript, we employ first-principles calculations to study the spin-valley physics of \MS/\WS~heterostructures, focusing on the high-symmetry stackings with 0\degree (R-type) and 60\degree (H-type) twist angle under the influence of external electric fields and variations of the interlayer distance. We perform a systematic analysis of the spin ($S_z$) and orbital ($L_z$) angular momenta, g-factors, and symmetry properties of the relevant band edges (at K, Q, and $\Gamma$ points) and reveal that different stackings imprint distinct hybridization signatures on the spin-valley physics of these low-energy bands. We follow the successful framework to compute the valley Zeeman splitting in TMDC materials\cite{Wozniak2020PRB, Deilmann2020PRL, Forste2020NatComm, Xuan2020PRR}, in which atomistic and ``valley'' contributions are correctly taken into account via the orbital angular momentum of the Bloch function. We extend our analysis to investigate direct- and indirect-in-momentum dipolar (interlayer) excitons, revealing that different exciton species present distinct electric dipole moments and polarizabilities, and g-factors. Particularly, direct dipolar excitons at the K-valley hold a robust valley Zeeman splitting nearly independent of the electric field, whereas the momentum-indirect dipolar excitons, with conduction bands at the Q point, show a clear dependence of the valley Zeeman splitting for positive values of applied electric fields (with variations of $\sim4$). The dipolar exciton g-factors can also be affected by varying the interlayer distance, which could be experimentally probed under external pressure. Because of the opposite orientation of $S_z$ and $L_z$ of the individual monolayers, our results reveal that changes in the valley Zeeman due to electric field or interlayer distance are more pronounced in H-type systems, such as the g-factor variation from $-5$ to $3$ in the valence bands of the \Hh~stacking. These pronounced valley Zeeman signatures should be experimentally accessible in samples with a twist angle close to 60$\degree$, which give rise to moiré or atomic reconstructed domains dominated by the high-symmetry H-type stackings.

This manuscript is organized as follows: In Section~\ref{sec:gen}~we discuss the electronic structure and symmetry properties of the studied R- and H-type high-symmetry stackings. In Section~\ref{sec:spinvalley}, we explore the spin-valley physics of the relevant band edges and how they are affected by an applied electric field and variations of the interlayer distance. In Section~\ref{sec:inter_exc}, we focus on the direct and indirect, in $k$-space, dipolar excitons originated from the low-energy band edges. We discuss their selection rules based on the symmetry analysis and first-principles calculations, their valley Zeeman signatures (effective g-factors), and how the effect of external electric fields and variations of the interlayer distance impact the optical selection rules and valley Zeeman physics. Finally, in Section~\ref{sec:conclusions}, we summarize our findings and place our study on the broader context of the growing research field of valleytronics in multilayered van der Waals heterostructures.

%================================================================================

\section{General features of the M\lowercase{o}S\lowercase{e}$_2$/WS\lowercase{e}$_2$ high-symmetry stackings}
\label{sec:gen}

We focus on artificially stacked \MS/\WS~heterostructures with relative angles of 0\degree~(R-type) and 60\degree~(H-type). Certainly, when the individual \MS~and \WS~layers are stacked together, the angle is not precisely 0\degree~or 60\degree~and the small misalignment can give rise to moiré superlattices or atomic reconstruction\cite{Tran2019Nature, Seyler2019Nature, Choi2021PRL, Enaldiev2020PRL, Forg2021NatComm,  Holler2020APL, Parzefall2021TDM, Zhao2022} with length scales of tens of nm. However, either in the moiré or in the atomic reconstruction perspective, there are distinct contributions from well defined high-symmetry stackings that dominate the observed properties\cite{Yu2015PRL, Yu2017SciAdv, Seyler2019Nature, Wozniak2020PRB, Nieken2022APLmat}, such as the dipolar excitons selection rules and g-factors. Furthermore, these high-symmetry stackings also serve as the building blocks for the effective modelling of interlayer moiré excitons\cite{Yu2015PRL, Yu2017SciAdv, Wu2017PRL, Yu2018TDM, Wu2018PRB, RuizTijerina2020PRB, Erkensten2021PRB}.

\begin{figure*}
\begin{center}
\includegraphics{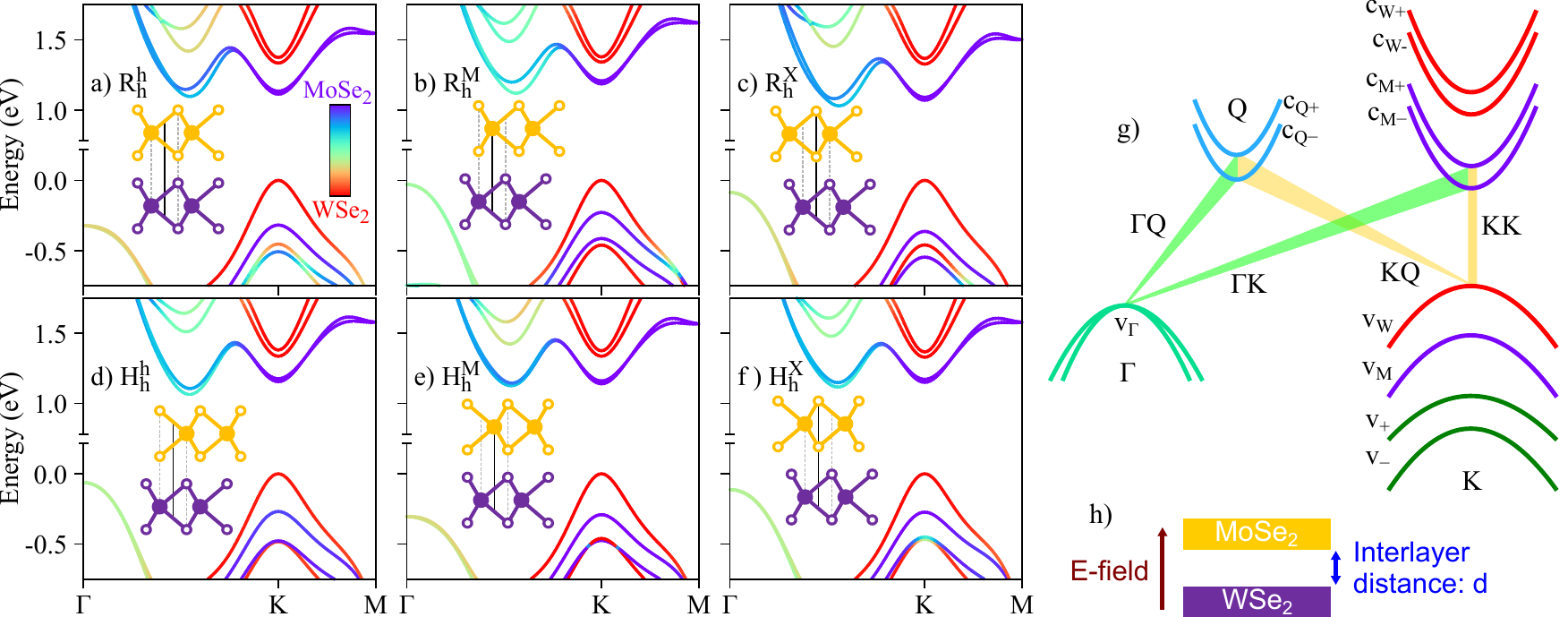}
\end{center}
\caption{Calculated band structures with the color-coded layer decomposition of the wave function for a) \Rh, b) \RM, c) \RX, d) \Hh, e) \HM, and f) \HX~stackings. The insets show the side view of the crystal structures (solid lines connect the hollow position of the bottom layer to the atomic registry of the top layer). g) Relevant low-energy bands and possible interlayer exciton transitions originating from the top valence bands at the $\Gamma$ (v$_\Gamma$) and K points (v$_{\textrm{W}}$). The transitions involving the time-reversal partners ($-$K and $-$Q points) are not shown for simplicity. h) Schematic representation of the applied external electric field and the interlayer distance.}
\label{fig:bs_qtl_GKM_scheme}
\end{figure*}

The side view of the considered high-symmetry stackings (\Rh, \RM, \RX, \Hh, \HM~and \HX) are shown as insets in Figs.~\ref{fig:bs_qtl_GKM_scheme}a-f and we follow the labeling of Refs.~\cite{Tran2019Nature,Wozniak2020PRB}. To clarify, the stacking label X$^{\textrm{a}}_{\textrm{b}}$ indicates the twist angle X = R (0\degree) or X = H (60\degree) and relative shifts between bottom (subscript a) and top (superscript b) layer, with a,b = h (hollow), M (metal), and X (chalcogen). For instance, the \RX~stacking indicates a structure with 0\degree~with the chalcogen atom of the top layer positioned on top of the hollow position of the bottom layer. To construct the heterostructure, we start with the values of in-plane lattice parameter ($a$) and thickness ($t$) of the individual monolayers given by $a=3.289\;\textrm{\AA}$, $t=3.335\;\textrm{\AA}$ for \MS, $a=3.282\;\textrm{\AA}$, $t=3.340\;\textrm{\AA}$ for \WS, taken from Ref.~\cite{Kormanyos2015TDM}. For the \MS/\WS~heterostructures, we consider an average lattice parameter of $a=3.2855\;\textrm{\AA}$ leading to an average strain of $\sim 0.1 \%$.  We also correct the monolayer thickness due to strain\cite{Zollner2019PRB}, leading to $t=3.3375\;\textrm{\AA}$ and $t=3.3374\;\textrm{\AA}$ for \MS~and \WS, respectively. A vacuum region of $20\;\textrm{\AA}$ is considered for all cases. The equilibrium interlayer distances, $d$, are given in Table~\ref{tab:ix} and are consistent with several reports in the literature\cite{Nayak2017ACSNano, Gillen2018PRB, Torun2018PRB, Xu2018PCCP, Tran2019Nature, Wozniak2020PRB, Gillen2021pssb}. We also present in Table~\ref{tab:ix} the effective distance between Mo and W atoms. We performed our first-principles calculations based on the density functional theory (DFT) as implemented in the WIEN2k\cite{wien2k} code, with details given in Appendix A.

\begin{table}
\caption{Interlayer distance, d, and effective Mo-W distance for the studied R- and H-type systems.}
\begin{center}
\begin{tabular}{ccccccc}
\hline
\hline
 & \Rh & \RM & \RX & \Hh & \HM & \HX  \\ [1mm]
 \hline
d ($\textrm{\AA}$) & 3.7237 & 3.0803 & 3.0869 & 3.0923 & 3.6885 & 3.1833 \\
Mo-W ($\textrm{\AA}$) & 7.0612 & 6.6922 & 6.6985 & 6.7037 & 7.2775 & 6.5208 \\
\hline
\hline
\end{tabular}
\end{center}
\label{tab:ix}
\end{table}

\begin{figure*}
\begin{center} 
\includegraphics{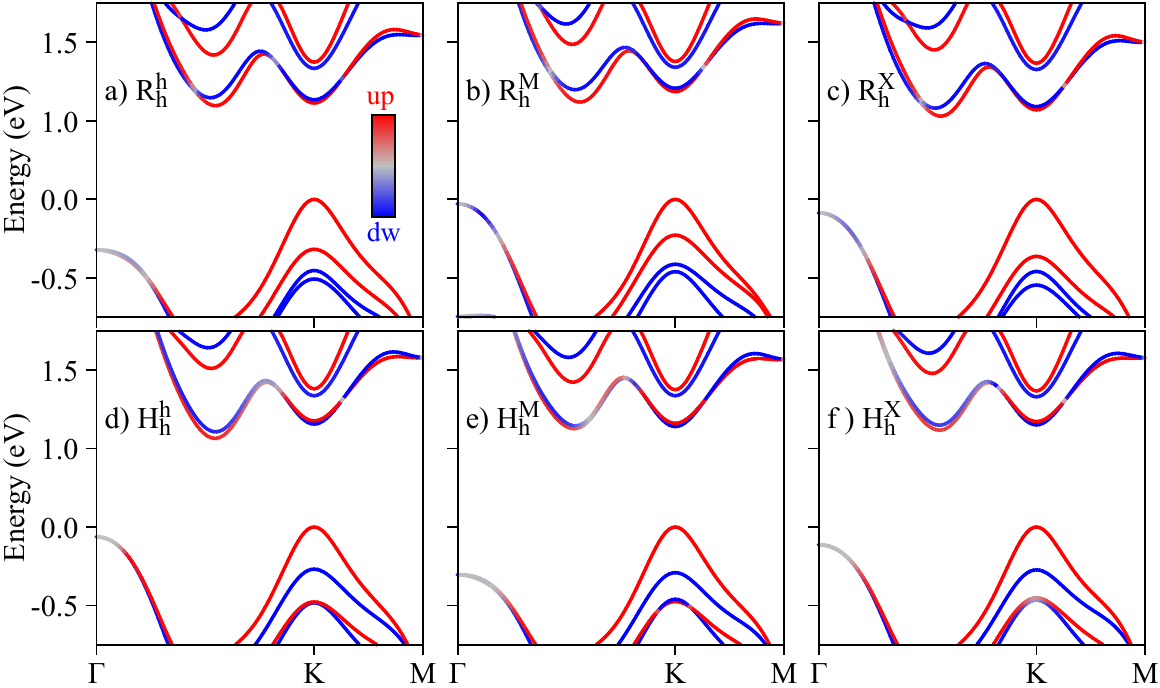}
\end{center}
\caption{Spin-resolved band structures for the a) \Rh, b) \RM, c) \RX, d) \Hh, e) \HM~and f) \HX~stackings.}
\label{fig:bs_spin_GKM}
\end{figure*}

The layer-resolved band structures for all the calculated R- and H-type structures are shown in Figs.~\ref{fig:bs_qtl_GKM_scheme}a-f. We are particularly interested in the band edges that can host dipolar excitons (K, Q, and $\Gamma$ points) as depicted in Fig.~\ref{fig:bs_qtl_GKM_scheme}g, and how they are affected by external electric fields and changes of the interlayer distance, schematically shown in Fig~\ref{fig:bs_qtl_GKM_scheme}h. The conduction and top valence bands at the K point are highly layer-localized and maintain the same order regardless of the stacking (bands $\textrm{c}_{\textrm{W}\pm}$, $\textrm{c}_{\textrm{M}\pm}$, $\textrm{v}_{\textrm{W}}$, and $\textrm{v}_{\textrm{M}}$). On the other hand, the  ordering of the lowest valence bands depend on the particular stacking (we therefore just identify them as $\textrm{v}_+$ and $\textrm{v}_-$, specifying the majority layer composition when necessary). The energy bands outside of the K point often show a higher interlayer hybridization, as is the case of the low-energy conduction bands at the Q point ($\textrm{c}_{\textrm{Q}\pm}$) and the top valence band at the $\Gamma$ point ($\textrm{v}_\Gamma$). Because of the different twist angle (0\degree or 60\degree) between the layers, the relative spin degree of freedom is certainly affected, as indicated by the spin-resolved band structure shown in Fig.~\ref{fig:bs_spin_GKM}.

\begin{table}
\caption{Irreducible representations at the K point ($C_3$ point group) for the relevant energy bands indicated in Fig.~\ref{fig:bs_qtl_GKM_scheme}g for all considered stackings.}
\begin{center}
\begin{tabular}{lcccccc}
\hline
\hline
 & \Rh & \RM & \RX & \Hh & \HM & \HX  \\ [1mm]
 \hline
$\textrm{c}_{\textrm{W}+}$ & $K_{5}$ & $K_{6}$ & $K_{4}$ & $K_{4}$ & $K_{6}$ & $K_{5}$ \\
$\textrm{c}_{\textrm{W}-}$ & $K_{4}$ & $K_{5}$ & $K_{6}$ & $K_{6}$ & $K_{5}$ & $K_{4}$ \\
$\textrm{c}_{\textrm{M}+}$ & $K_{4}$ & $K_{4}$ & $K_{4}$ & $K_{5}$ & $K_{5}$ & $K_{5}$ \\
$\textrm{c}_{\textrm{M}-}$ & $K_{5}$ & $K_{5}$ & $K_{5}$ & $K_{4}$ & $K_{4}$ & $K_{4}$ \\
\hline
$\textrm{v}_{\textrm{W}}$ & $K_{4}$ & $K_{5}$ & $K_{6}$ & $K_{6}$ & $K_{5}$ & $K_{4}$ \\
$\textrm{v}_{\textrm{M}}$ & $K_{4}$ & $K_{4}$ & $K_{4}$ & $K_{5}$ & $K_{5}$ & $K_{5}$ \\
$\textrm{v}_{+}$ & $K_{6}$ & $K_{6}$ & $K_{5}$ & $K_{6}$ & $K_{4}$ & $K_{6}$ \\
$\textrm{v}_{-}$ & $K_{6}$ & $K_{4}$ & $K_{6}$ & $K_{5}$ & $K_{6}$ & $K_{6}$ \\
\hline
\hline
\end{tabular}
\end{center}
\label{tab:irreps}
\end{table}

In terms of symmetry, the R- and H-type structures considered belong to the $C_{3v}$ (P3m1) point group. In reciprocal space, the symmetry group of the $\Gamma$ point is also the $C_{3v}$ group and is reduced to its subgroups for other $k$-points. We find the $C_{3}$ group for the K point, the $C_1$ group for the Q point, and the $C_s$ group for the M point. It is also relevant to identify the double-group (with SOC) irreducible representations (irreps) for the relevant energy bands shown in Fig.~\ref{fig:bs_qtl_GKM_scheme}g. These irreps are useful building blocks to study the allowed optical selection rules (direct and phonon-mediated), and possible (intra- and/or inter-valley) scattering mechanisms, as already studied in the monolayer cases\cite{Song2013PRL, Dery2015PRB, Gilardoni2021PRB, Blundo2022PRL, Smirnov2022TDM}. Furthermore, irreps are crucial to build effective models within the \textbf{k.p} framework\cite{Dresselhaus1955PR, Kane1966, Kane1957JPCS, dresselhaus, voon, Kormanyos2015TDM}, in which additional symmetry-breaking terms can be easily incorporated. Specifically for the K point, the irreps of the energy bands are given in Table~\ref{tab:irreps}. Since the $K_{4}$ and $K_{5}$ irreps are complex, the irreps at $-$K can be found by taking the complex conjugate $K_{4} \rightarrow K_{5}=K^*_{4}$, $K_{5} \rightarrow K_{4}=K^*_{5}$. The top valence band at the $\Gamma$ point (2-fold degenerate) belongs to the (real) irrep $\Gamma_{4}$ of the $C_{3v}$ group and the lower conduction bands at the Q point both belong to the (real) irrep $Q_2$ of the $C_1$ group. To clarify our notation, we identify the irreps by their reciprocal space point, i. e., K$_i$ irreps belong to the K point, $\Gamma_i$ irreps for the $\Gamma$ point and Q$_i$ irreps for the Q point. We also emphasize that the irreps are obtained using the full wave function calculated within DFT, as implemented in WIEN2k\cite{wien2k}. All the irreps and symmetry groups discussed here follow the character tables of Ref.~\cite{koster}.

%================================================================================

\section{Spin-valley physics at the band edges}
\label{sec:spinvalley}

%--------------------------------------------------------------------------------

\subsection{Pristine heterostructures}

\begin{table}
\caption{Calculated values of $S_z$ and $L_z$ for the relevant energy bands in monolayers.}
\begin{center}
\begin{tabular}{crrrr}
\hline
\hline
 & \multicolumn{2}{c}{$\quad$\MS} & \multicolumn{2}{c}{$\quad$\WS} \\
 & $S_{z}$ & $L_{z}$ & $S_{z}$ & $L_{z}$ \\
 \hline
$\textrm{c}_{+}$ & $-0.99$ & $1.53$ & $0.99$ & $2.98$ \\
$\textrm{c}_{-}$ & $1.00$ & $1.81$ & $-0.89$ & $1.88$ \\
\hline
$\textrm{v}_{+}$ & $1.00$ & $3.94$ & $1.00$ & $5.02$ \\
$\textrm{v}_{-}$ & $-0.99$ & $3.67$ & $-0.98$ & $4.07$ \\
\hline
$\textrm{v}_{\Gamma}$ & $0.96$ & $0.06$ & $0.87$ & $0.19$ \\
$\textrm{c}_{\textrm{Q}+}$ & $-1.00$ & $-0.09$ & $-0.99$ & $0.32$ \\
$\textrm{c}_{\textrm{Q}-}$ & $1.00$ & $-0.16$ & $0.96$ & $0.05$ \\
\hline
\hline
\end{tabular}
\end{center}
\label{tab:SLmono}
\end{table}

\begin{table*}[htb]
\caption{Calculated values of $S_z$ and $L_z$ for the relevant energy bands in R- and H- type stackings.}
\begin{center}
\begin{tabular}{crrrrrrrrrrrr}
\hline
\hline
 & \multicolumn{2}{c}{$\quad$\Rh} & \multicolumn{2}{c}{$\quad$\RM} & \multicolumn{2}{c}{$\quad$\RX} & \multicolumn{2}{c}{$\quad$\Hh} & \multicolumn{2}{c}{$\quad$\HM} & \multicolumn{2}{c}{$\quad$\HX} \\
 & $S_{z}$ & $L_{z}$ & $S_{z}$ & $L_{z}$ & $S_{z}$ & $L_{z}$ & $S_{z}$ & $L_{z}$ & $S_{z}$ & $L_{z}$ & $S_{z}$ & $L_{z}$ \\
 \hline
$\textrm{c}_{\textrm{W}+}$ & $0.97$ & $2.98$ & $0.97$ & $3.02$ & $0.97$ & $2.97$ & $0.97$ & $2.92$ & $0.97$ & $2.98$ & $0.97$ & $3.02$ \\
$\textrm{c}_{\textrm{W}-}$ & $-0.88$ & $1.88$ & $-0.87$ & $1.93$ & $-0.88$ & $1.88$ & $-0.85$ & $1.86$ & $-0.89$ & $1.89$ & $-0.88$ & $1.90$ \\
$\textrm{c}_{\textrm{M}+}$ & $-0.98$ & $1.53$ & $-0.98$ & $1.55$ & $-0.98$ & $1.54$ & $0.98$ & $-1.54$ & $0.98$ & $-1.52$ & $0.98$ & $-1.56$ \\
$\textrm{c}_{\textrm{M}-}$ & $1.00$ & $1.81$ & $0.99$ & $1.84$ & $1.00$ & $1.81$ & $-1.00$ & $-1.80$ & $-1.00$ & $-1.80$ & $-1.00$ & $-1.81$ \\
\hline
$\textrm{v}_{\textrm{W}}$ & $1.00$ & $5.02$ & $1.00$ & $4.99$ & $1.00$ & $4.98$ & $1.00$ & $4.79$ & $1.00$ & $5.01$ & $1.00$ & $5.02$ \\
$\textrm{v}_{\textrm{M}}$ & $1.00$ & $3.94$ & $0.99$ & $3.94$ & $1.00$ & $3.91$ & $-1.00$ & $-3.25$ & $-1.00$ & $-3.93$ & $-1.00$ & $-3.95$ \\
$\textrm{v}_{+}$ & $-0.98$ & $4.04$ & $-0.99$ & $3.63$ & $-0.98$ & $4.01$ & $0.99$ & $-3.45$ & $-0.98$ & $4.08$ & $0.24$ & $-0.67$ \\
$\textrm{v}_{-}$ & $-0.99$ & $3.70$ & $-0.96$ & $4.02$ & $-0.99$ & $3.62$ & $-0.98$ & $3.41$ & $0.99$ & $-3.66$ & $-0.22$ & $1.06$ \\
\hline
$\textrm{v}_{\Gamma}$ & $0.92$ & $0.09$ & $0.95$ & $0.04$ & $0.95$ & $0.05$ & $0.95$ & $0.05$ & $0.93$ & $0.08$ & $0.95$ & $0.05$ \\
$\textrm{c}_{\textrm{Q}+}$ & $-0.93$ & $0.11$ & $-0.91$ & $0.23$ & $-0.87$ & $0.14$ & $-0.80$ & $0.22$ & $-0.59$ & $0.12$ & $-0.69$ & $0.23$ \\
$\textrm{c}_{\textrm{Q}-}$ & $0.95$ & $0.06$ & $0.92$ & $0.19$ & $0.92$ & $0.09$ & $0.79$ & $0.22$ & $0.60$ & $0.17$ & $0.68$ & $0.23$ \\
\hline
\hline
\end{tabular}
\end{center}
\label{tab:SLhet}
\end{table*}

In this section we explore the spin and orbital degrees of freedom for the relevant band edges (indicated in Fig.~\ref{fig:bs_qtl_GKM_scheme}g) at zero electric field and at the equilibrium interlayer distance. We follow the recent first-principles developments to compute the orbital angular momenta in monolayer TMDCs\cite{Wozniak2020PRB, Deilmann2020PRL, Forste2020NatComm, Xuan2020PRR}, which has also been successfully applied to investigate a variety of more complex TMDC-based systems and van der Waals heterostructures\cite{Xuan2021npj, Gillen2021pssb, Heissenbuettel2021NL, Forg2021NatComm, Covre2022Nanoscale, Blundo2022PRL, FariaJunior2022NJP, Raiber2022NatComm, Amit2022, Hotger2022, Gobato2022NL, Zhao2022, FariaJunior2023}. We do not aim to provide a detailed description of the methodology here, but briefly summarize the main points. An external magnetic field in the out-of-plane ($z$) direction (same orientation as the electric field in Fig.~\ref{fig:bs_qtl_GKM_scheme}h) modifies the energy levels as
\begin{align}
E_{\textrm{ZS}}(n,\vec{k}) & = g_z(n,\vec{k})\mu_{B}B \nonumber \\ 
& = \left[S_z(n,\vec{k})+L_z(n,\vec{k})\right]\mu_{B}B  \, ,
\label{eq:EZ}
\end{align}
in which the g-factor, $g_z(n,\vec{k})$, of the Bloch band (labeled by $n$ and $\vec{k}$) is written in terms of the spin and orbital angular momenta, given by
\begin{align}
S_z(n,\vec{k}) & = \left\langle n,\vec{k}\,\bigl|\sigma_{z}\bigr|n,\vec{k}\right\rangle \nonumber \\
L_z(n,\vec{k}) & = \frac{1}{im_{0}} \underset{m\neq n}{\sum} \! \frac{P_{x}^{n,m,\vec{k}}P_{y}^{m,n,\vec{k}} \! - \! P_{y}^{n,m,\vec{k}}P_{x}^{m,n,\vec{k}}}{E(n,\vec{k})-E(m,\vec{k})} \, ,
\label{eq:LzSz}
\end{align}
in which $\sigma_z$ is the Pauli matrix for the spin components along $z$, $P_{\alpha}^{n,m,\vec{k}}=\left\langle n,\vec{k}\,\bigl| {p}_{\alpha}\bigr| m,\vec{k} \right\rangle$ ($\alpha=x,y,z$), $\vec{p}$ is the momentum operator, $E(n,\vec{k})$ is the energy of the Bloch band and $m_0$ is the free electron mass. In the expression of the spin angular momentum, we have used $g_{0} \approx 2$ for the electron spin g-factor. The summation-over-bands that characterizes $L_z$ requires a sufficiently large number of bands ($\sim$500) to achieve convergence, which is properly taken into account in our DFT calculations (see details in Appendix A). These \MS/\WS~heterostructures have been investigated in Refs.~\cite{Wozniak2020PRB, Xuan2020PRR, Gillen2021pssb, Forg2021NatComm, Zhao2022} but here we extend our analysis to incorporate more bands and later investigate the dependence of electric field and interlayer distance. We emphasize here that the electronic and spin properties calculated with the DFT also provide a reliable benchmark to further studies since WIEN2k\cite{wien2k} is one of the most accurate DFT codes available\cite{Lejaeghere2016Science}) and particularly suitable to study spin-physics of 2D materials and van der Waals heterostructures\cite{Gmitra2009PRB, Konschuh2010PRB, Gmitra2015PRB, Gmitra2016PRB, Kurpas2016PRB, Kurpas2019PRB, FariaJunior2019PRB, FariaJunior2022NJP, FariaJunior2023}.

Our calculated values of $S_z$ and $L_z$ for the monolayers are collected in Table~\ref{tab:SLmono} and the results for the R- and H-type stackings are shown in Table~\ref{tab:SLhet}. For the monolayer cases we used the same in-plane lattice parameters and thicknesses as used in the heterostructure so that we can directly compare the monolayer values of $S_z$ and $L_z$ to the heterostructure. The amount of strain of $\sim 0.1 \%$ for the in-plane lattice parameter has a very small influence on the spin-valley physics of individual monolayer TMDCs\cite{Zollner2019PRB, FariaJunior2022NJP}, with contributions on the order of $10^{-2}$ from the unstrained values. Therefore, any changes of $L_z > 10^{-2}$ are clear signatures of interlayer hybridization. For the heterostructure bands at the K point, $\textrm{c}_{\textrm{W(Mo)}\pm}$, $\textrm{v}_{\textrm{W(Mo)}}$ and $\textrm{v}_{\pm}$, we observe the typical sign changes from R to H stackings because of the opposite K/K ($0^\circ$ twist) and K/-K ($60^\circ$ twist) alignment\cite{Seyler2019Nature, Wozniak2020PRB}. For the conduction bands, we observe deviations of $L_z$ on the order of $\leq 0.05$ when compared to the monolayer case, consistent with a higher localization of conduction bands\cite{FariaJunior2023}. We therefore would expect larger changes arising in the valence bands, which is indeed the case, particularly for the $\textrm{v}_{\pm}$ bands, and for the $\textrm{v}_{\textrm{W}}$ and $\textrm{v}_{\textrm{Mo}}$ bands of the \Hh~stacking. Furthermore, the effects of band mixing in the R stacking are less pronounced because at zero field and at the equilibrium distance, the values of $L_z$ have the same sign and similar magnitude, whereas in the H stacking $L_z$ has opposite sign and any admixture of the energy bands can significantly alter $L_z$. For the highly delocalized bands at $\Gamma$ and Q points, we found values of $S_z$ and $L_z$ that are in between the values of individual \MS~and \WS monolayers, but still with $\left| S_z \right| > \left| L_z \right|$, thus making spin the dominant degree of freedom, as in the monolayer case\cite{FariaJunior2022NJP}. The g-factor values, $g_z$, can be found by adding $S_z$ and $L_z$ (we also present $g_z$ graphically in the Sections~\ref{sec:bands_Efield} and \ref{sec:bands_iX}, as function of the electric field and interlayer distance).

%--------------------------------------------------------------------------------

\subsection{Electric field dependence}
\label{sec:bands_Efield}

\begin{figure*}
\begin{center}
\includegraphics{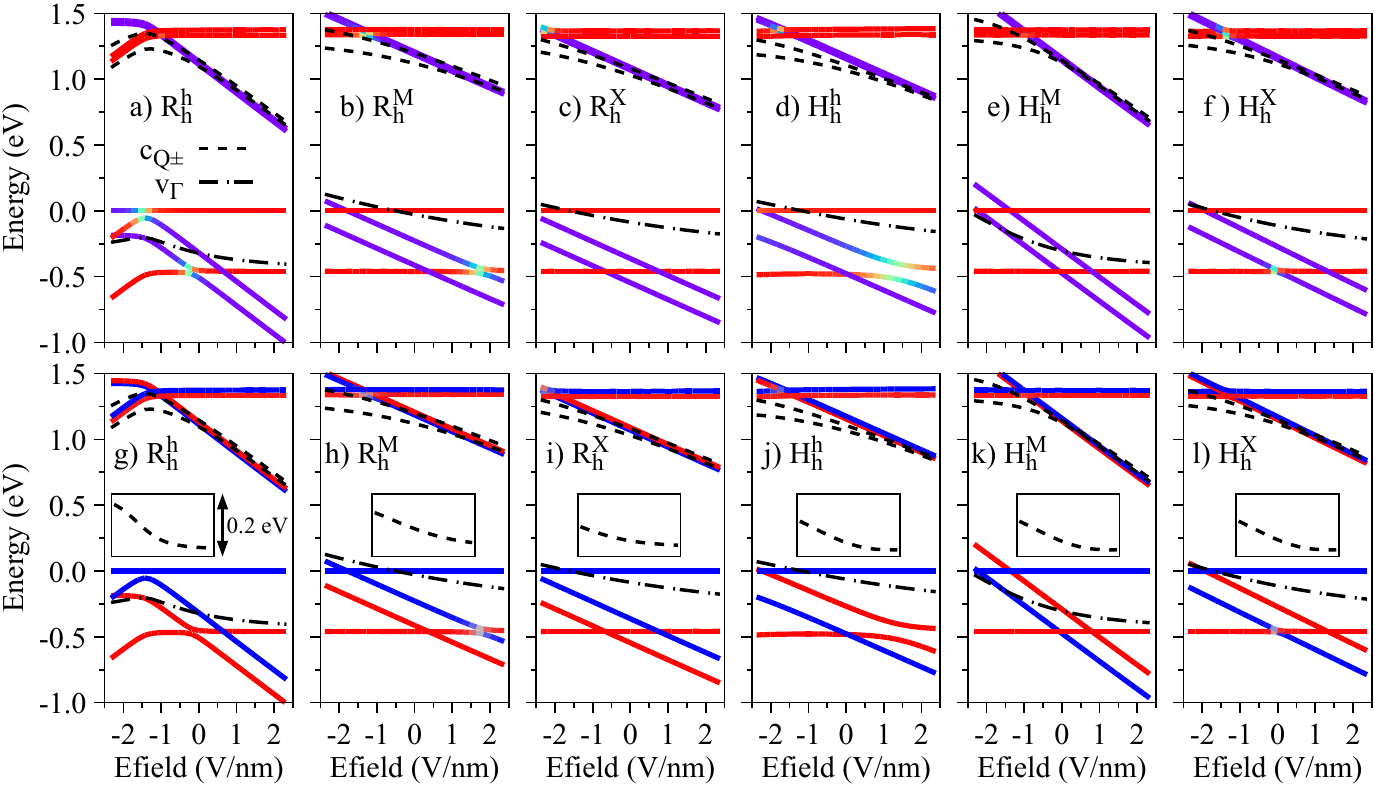}
\end{center}
\caption{Energy dependence with respect to the electric field of the relevant low-energy bands (see Fig.~\ref{fig:bs_qtl_GKM_scheme}g) for all considered stackings. The top row, panels~a-f, indicates the color-coded layer decomposition of the K-point bands (same color code as in Figs.~\ref{fig:bs_qtl_GKM_scheme}a-f). The bottom row, panels~g-l, indicates the color-coded spin decomposition of the K-point bands (same color code as in Figs.~\ref{fig:bs_spin_GKM}a-f). The insets in panels~g-l show the energy difference between $\textrm{c}_{\textrm{Q}+}$ and $\textrm{c}_{\textrm{Q}-}$ bands emphasizing an anti-crossing at larger electric fields.}
\label{fig:Efield_qtl_spin_line}
\end{figure*}

\begin{figure*}
\begin{center}
\includegraphics{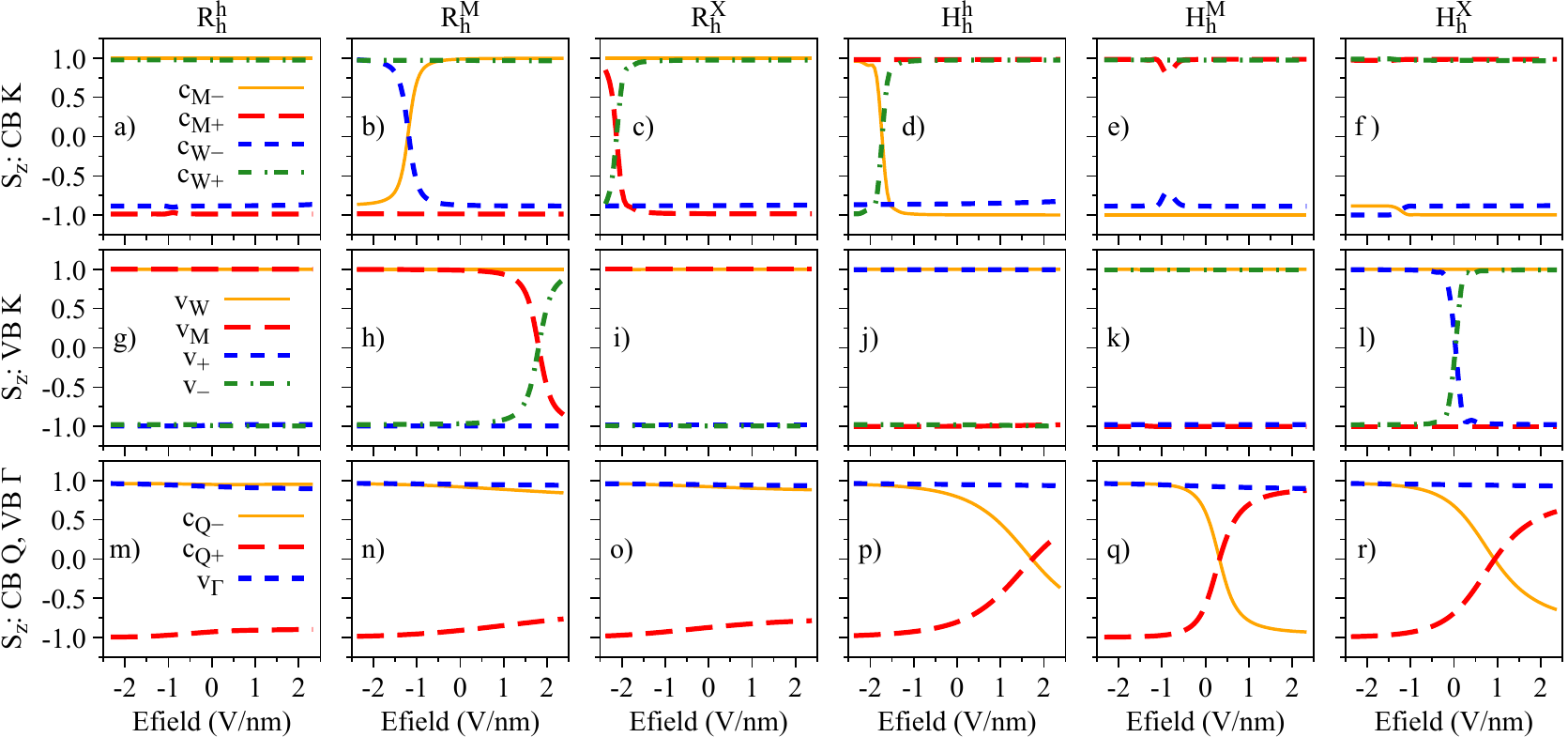}
\end{center}
\caption{Spin degree of freedom, $S_z$, of the low-energy bands as a function of the applied electric field for the studied stackings \Rh~(a,g,m), \RM~(b,h,n), \RX~(c,i,o), \Hh~(d,j,p), \HM~(e,k,q) and \HX~(f,l,r).}
\label{fig:Efield_Sz}
\end{figure*}

We now investigate the effect of external electric fields on the band edges of the R- and H-type stackings. In Fig.~\ref{fig:Efield_qtl_spin_line} we show the energy dependence of the relevant energy bands (see Fig.~\ref{fig:bs_qtl_GKM_scheme}g) as function of the electric field with the color code of the K bands representing the layer-decomposition in the top row and the spin degree of freedom in the bottom row. Due to the particular symmetry of the energy bands (orbital and spin characters), the electric field can induce crossings or anti-crossings. The anti-crossings, visible in the top row of Fig.~\ref{fig:Efield_qtl_spin_line} when the color code deviates from red (\WS~layer) or blue (\MS~layer), indicate regions with strong interlayer hybridization. Furthermore, these anti-crossings can either mix or preserve the spin degrees of freedom, depending on the stacking and bands involved. For instance, for the \Hh~stacking there is an anti-crossing between $\textrm{v}_{\textrm{M}}$ and $\textrm{v}_{-}$ (mainly localized at the \WS~layer) that preserves spin. On the other hand, the anti-crossing of the \RM~stacking between $\textrm{v}_{\textrm{M}}$ and $\textrm{v}_{-}$ (mainly localized at the \WS~layer) also acts on the spin degree of freedom. We point out that the conduction bands at the Q point, shown by short dashed lines, do not present any crossing when the energy levels get closer as the values of electric field increase, as shown in the insets of Figs.~\ref{fig:Efield_qtl_spin_line}g-l. The crossing of conduction bands takes place when electric field values are smaller than $\sim-0.75$ V/nm.

\begin{figure*}
\begin{center}
\includegraphics{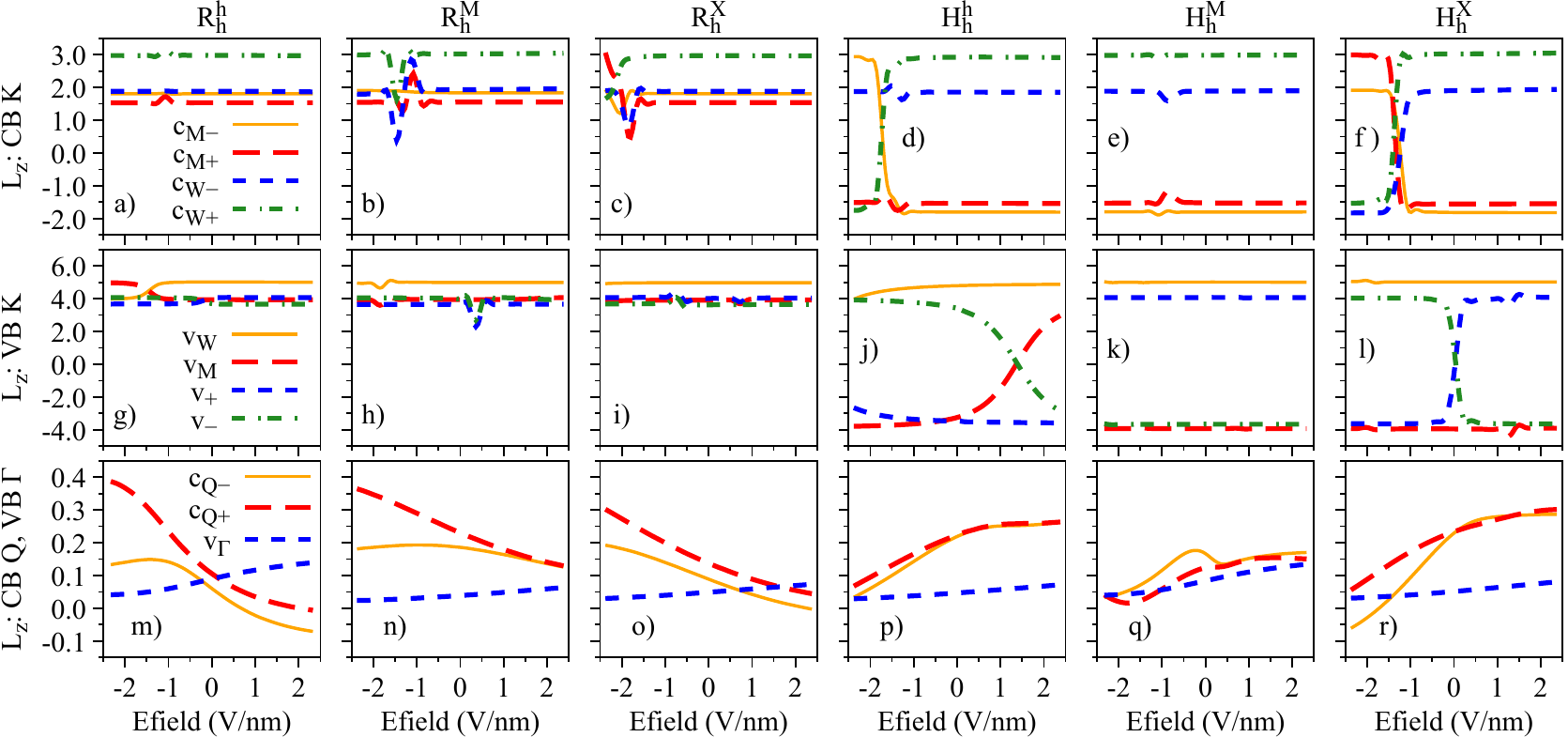}
\end{center}
\caption{Same as Fig.~\ref{fig:Efield_Sz} but for the orbital degree freedom, $L_z$.}
\label{fig:Efield_Lz}
\end{figure*}

\begin{figure*}
\begin{center}
\includegraphics{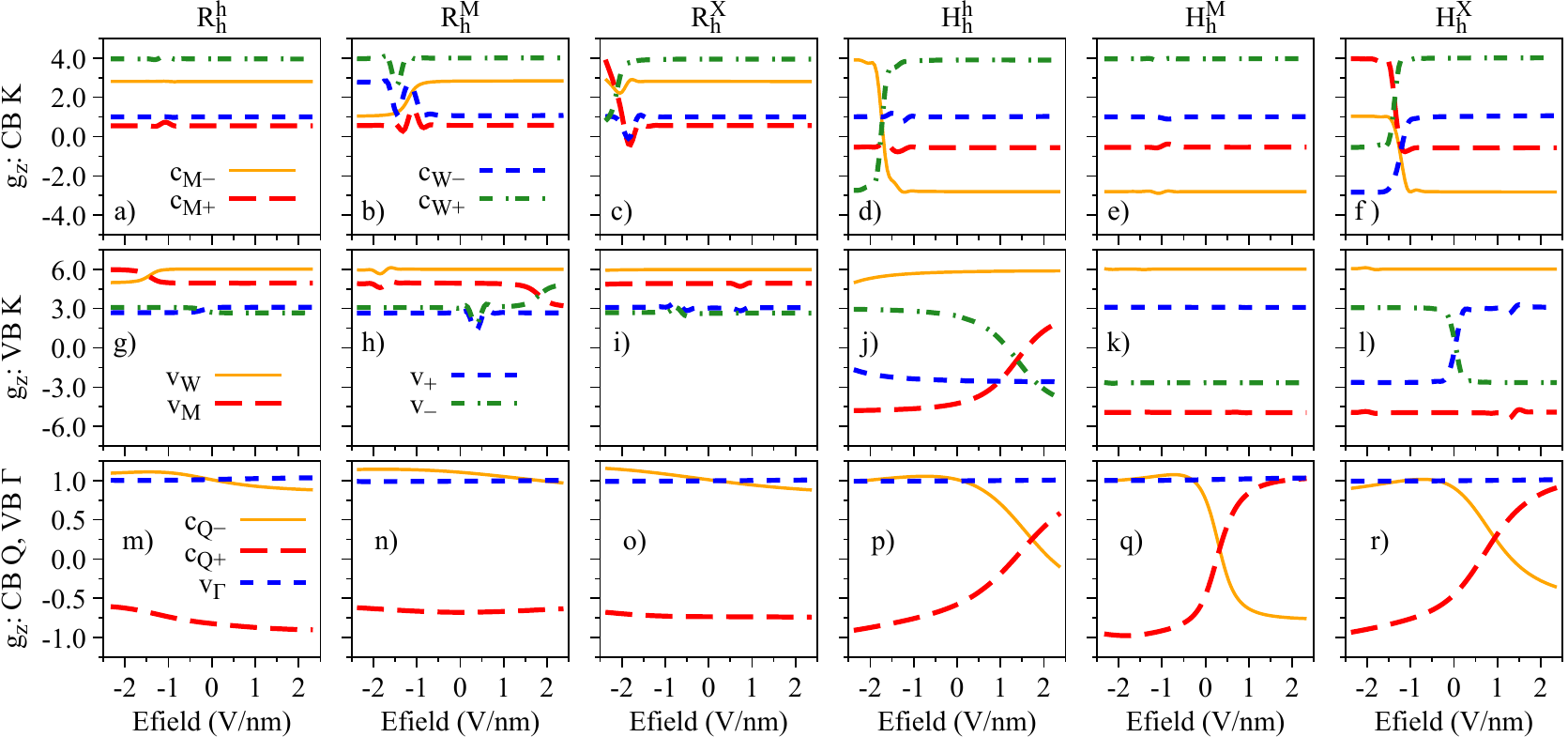}
\end{center}
\caption{Same as Figs.~\ref{fig:Efield_Sz} and \ref{fig:Efield_Lz} but for the band g-factor, $g_z = L_z + S_z$.}
\label{fig:Efield_gz}
\end{figure*}

The electric field dependence of the quantities $S_z$, $L_z$, and $g_z$ that encode the essence of spin-valley physics is summarized in Figs.~\ref{fig:Efield_Sz}, ~\ref{fig:Efield_Lz} and ~\ref{fig:Efield_gz}, respectively. The most important feature revealed by our calculations is that $S_z$, $L_z$, and $g_z$ are strongly mixed around the anti-crossing regions shown in Fig.~\ref{fig:Efield_qtl_spin_line}. In other words, spin and orbital degrees of freedom provide a clear quantitative signature for the strength of the interlayer coupling in these van der Waals heterostructures. Let us discuss in detail the case of \Hh~stacking, which shows the most prominent anti-crossing region for the valence bands $\textrm{v}_{\textrm{M}}$ and $\textrm{v}_{-}$. The orbital and spin analysis in Fig.~\ref{fig:Efield_qtl_spin_line} reveal that $\textrm{v}_{\textrm{M}}$ and $\textrm{v}_{-}$ mix on the orbital level while the spin remains unchanged. Indeed, by inspecting Fig.~\ref{fig:Efield_Sz}j, we confirm that $S_z$ is unchanged for the interval of applied electric field. On the other hand, Fig.~\ref{fig:Efield_Lz}j shows that the values of $L_z$ are drastically modified as the electric field swipes through the anti-crossing region, with the strongest mixing happening at $\sim1.5$ V/nm. In fact, already at zero electric field (as shown in Table~\ref{tab:SLhet}) we clearly observe the hybridization effect taking place in the values of $L_z$, suggesting that small values of applied electric field are enough to tune this hybridization even more. We point out that there are anti-crossing signatures also taking place in the conduction band. However, it seems that such feature has not been observed experimentally due to the range of electric field values applied\cite{Ciarrocchi2019NatPhot, Baek2020SciAdv, Barre2022Science}. Additionally, the electric field introduces a strong change of the spin character in the conduction bands at the Q point due to their anti-crossing (insets of Figs.~\ref{fig:Efield_qtl_spin_line}g-l), an interesting feature that can be investigated in low-energy interlayer excitons\cite{Barre2022Science}, which we discuss in further detail in Section~\ref{sec:inter_Efield}. We note that the field-driven Rashba SOC is naturally incorporated within our first principles calculations. However, the electric field is just too weak to effectively remove spins from their out-of-plane direction.

\begin{figure*}[h!]
\begin{center} 
\includegraphics{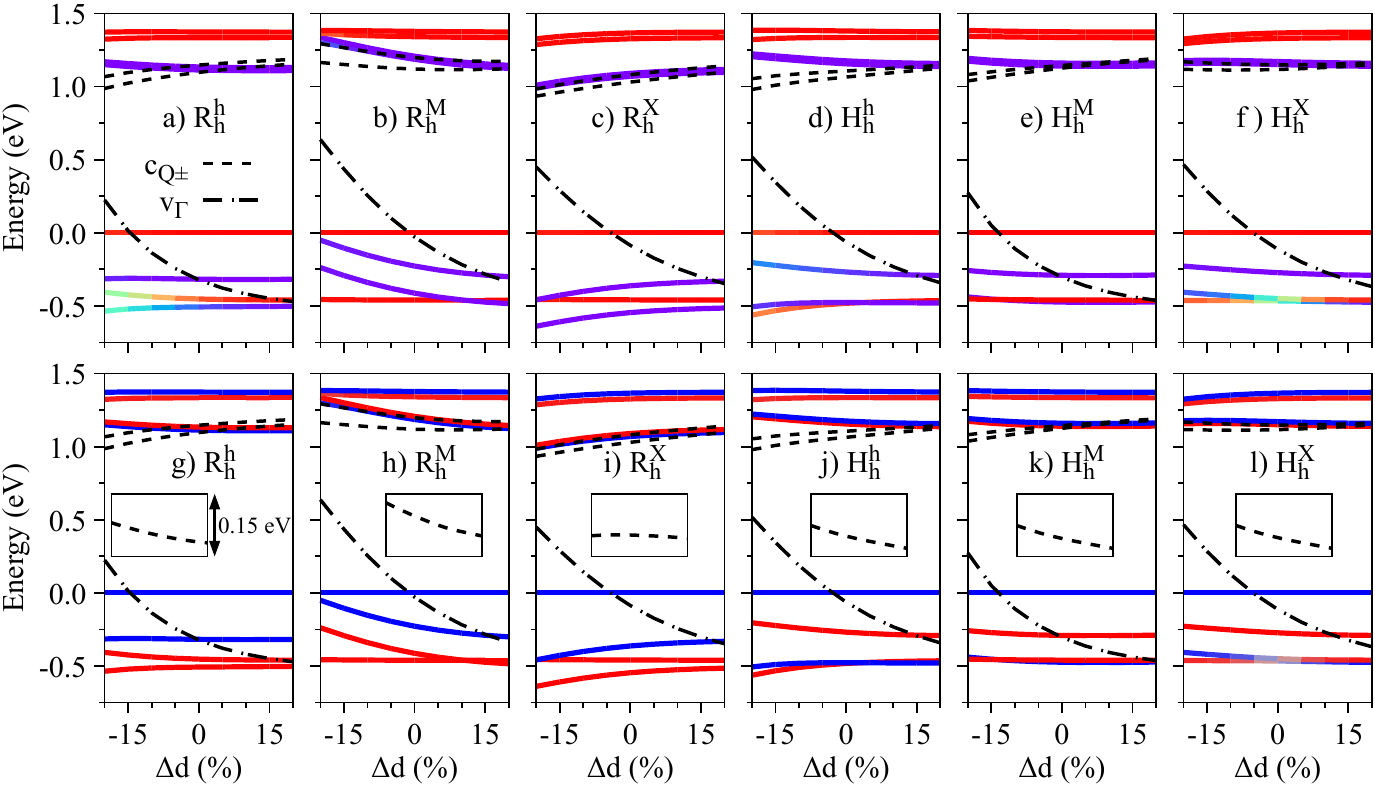}
\end{center}
\caption{Same as Fig.~\ref{fig:Efield_qtl_spin_line} but as a function of the interlayer distance variation.}
\label{fig:iX_qtl_spin_line}
\end{figure*}

\begin{figure*}
\begin{center}
\includegraphics{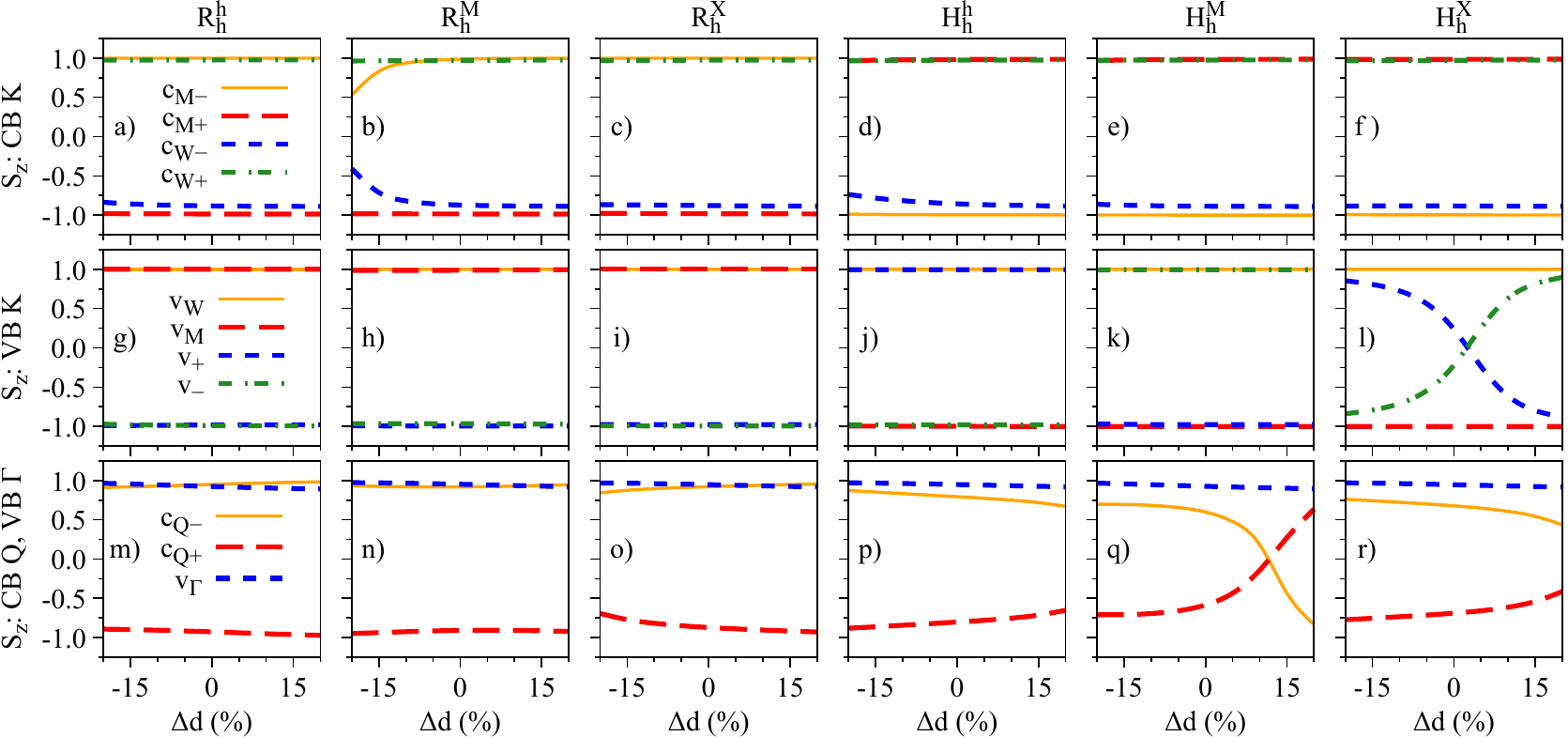}
\end{center}
\caption{Same as Fig.~\ref{fig:Efield_Sz} but as a function of the interlayer distance variation.}
\label{fig:iX_Sz}
\end{figure*}

Our detailed first-principles calculations reveal that the \Hh~stacking is the most suitable structure to investigate this interlayer hybridization signatures in the coupled spin-valley degrees of freedom. Surprisingly, this is the most stable structure for $60^\circ$ twist angles and the most relevant stacking for small twist angles, either having in mind a moiré or an atomically reconstructed sample, which facilitates the experimental accessibility to these effects. Furthermore, the opposite orientation of spin and orbital angular momenta from the opposite K-valleys in the individual monolayers favors g-factor variations from negative to positive values, for example, from $-3$ to $4$ in the conduction band (Fig.~\ref{fig:Efield_gz}d),  from $-5$ to $3$ in the valence band (Fig.~\ref{fig:Efield_gz}j) and from $-1$ to $1$ for conduction bands at the Q point (Fig.~\ref{fig:Efield_gz}p).

%--------------------------------------------------------------------------------

\subsection{Interlayer distance variation}
\label{sec:bands_iX}

In this section we investigate the consequences of varying the equilibrium interlayer distance. This can be particularly relevant in the context of fluctuations introduced by the mechanical exfoliation and stamping procedures in TMDC heterostructures\cite{Nagler2017NatComm, Plankl2021NatPhot, Parzefall2021TDM}, which creates regions with good and bad \textit{interlayer contact}. Additionally, changes in the interlayer distance can provide insights into the effects of external out-of-plane pressure, which can be currently investigated in available experimental setups, such as the recently reported studies on graphene/WSe$_2$\cite{Fulop2021npj2D}, multilayered TMDCs\cite{Oliva2022ACSami}, and WSe$_\text{2}$/Mo$_\text{0.5}$W$_\text{0.5}$Se$_\text{2}$ heterostructures\cite{Koo2023Light}. On a more abstract and theoretical level, artificially varying the interlayer distance provides an additional knob to modify the interlayer hybridization\cite{Kunstmann2018NatPhys} and analyze its impact on the spin and orbital angular momenta.

\begin{figure*}
\begin{center}
\includegraphics{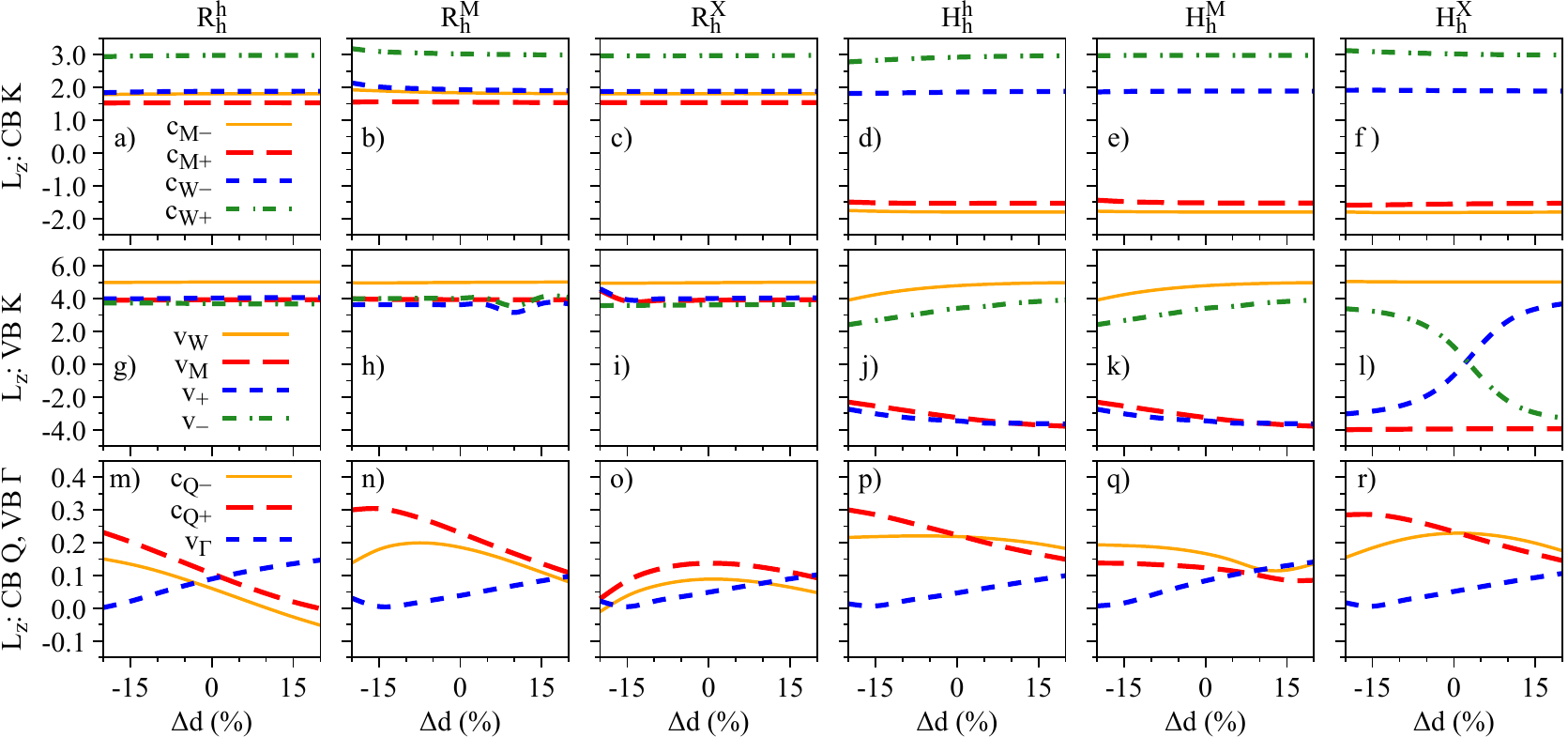}
\end{center}
\caption{Same as Fig.~\ref{fig:Efield_Lz} but as a function of the interlayer distance variation.}
\label{fig:iX_Lz}
\end{figure*}

\begin{figure*}
\begin{center}
\includegraphics{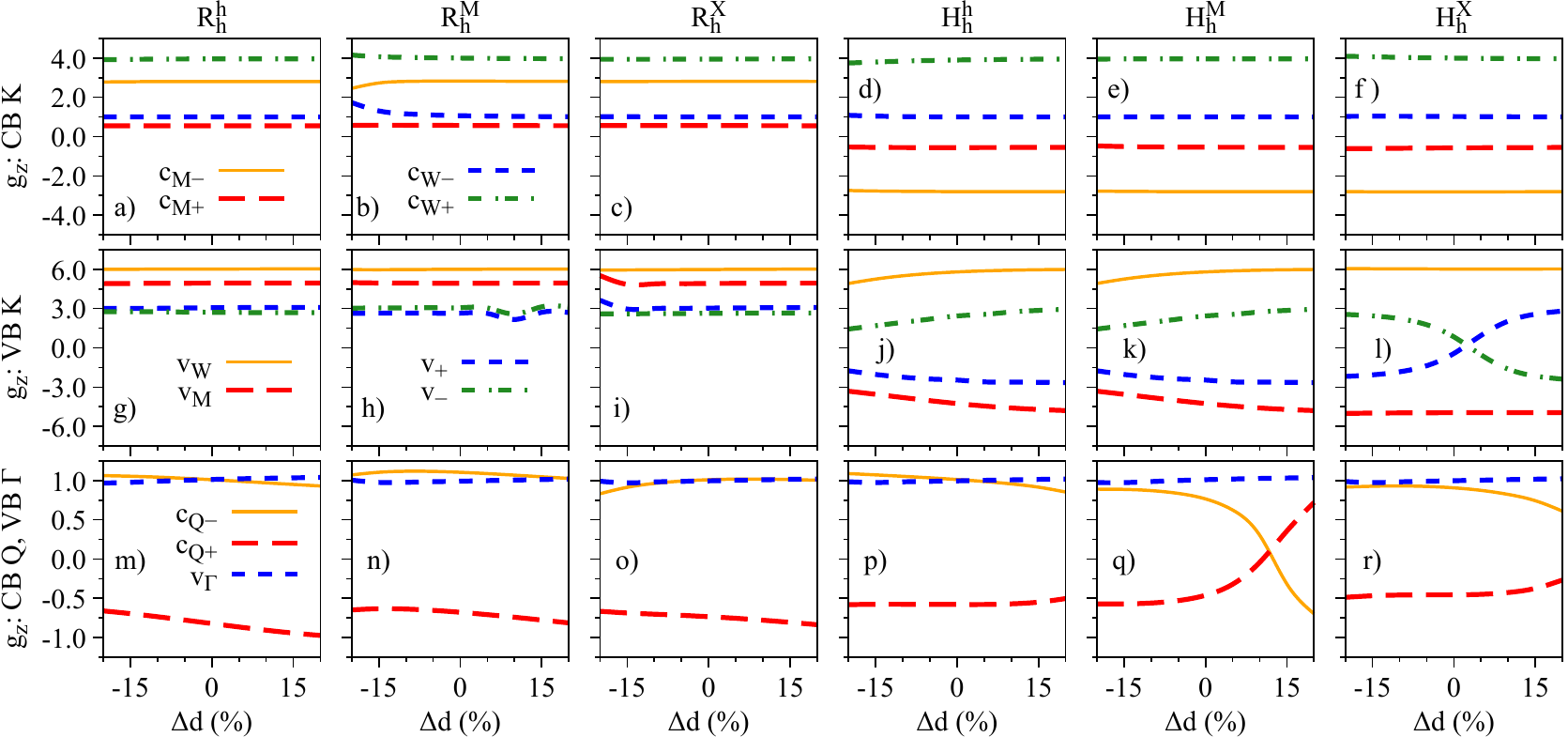}
\end{center}
\caption{Same as Fig.~\ref{fig:Efield_gz} but as a function of the interlayer distance variation.}
\label{fig:iX_gz}
\end{figure*}

The influence of the interlayer distance variation on the energy levels, spin angular momenta, orbital angular momenta, and g-factors is presented in Figs.~\ref{fig:iX_qtl_spin_line}, \ref{fig:iX_Sz}, \ref{fig:iX_Lz}, and \ref{fig:iX_gz}, respectively. We restrict ourselves to fluctuations of $\pm 20 \%$ of the equilibrium interlayer distances given in Table~\ref{tab:ix}. Overall, the same reasoning we discussed from the electric field effect applies to the interlayer distance variations. Nonetheless, we highlight a few points to strengthen the discussions and understanding presented in the previous Section~\ref{sec:bands_Efield}. There is a strong orbital mixing region visible in the lower valence bands with same spins in the \Rh~stacking case (Figs.~\ref{fig:iX_qtl_spin_line}a and \ref{fig:iX_qtl_spin_line}g), but with a negligible signature in the $L_z$ (Fig.~\ref{fig:iX_Lz}g) since the monolayer values of $L_z$ are all positive and close to each other. On the other hand, these hybridization effects are much stronger in the valence band of the \HX~stacking case (Figs.~\ref{fig:iX_qtl_spin_line}f and \ref{fig:iX_qtl_spin_line}l), with strong signatures in the values of $S_z$ and $L_z$ (Figs.~\ref{fig:iX_Sz}l and \ref{fig:iX_Lz}l). We also mention that in the \Hh~stacking, the most favorable for the 60\degree case, the $g_z$ values of the valence band (particularly $\textrm{v}_{\textrm{W}}$) can be effectively altered by reducing the interlayer distance (also seen in the \HM~stacking).

%================================================================================

\section{Low-energy dipolar excitons}
\label{sec:inter_exc}

In this section we investigate the fundamental ingredients that characterize the low-energy dipolar excitons, namely, the electric dipole moment, polarizability, dipole matrix elements and effective g-factors, based on the robust DFT calculations we presented in Sections~\ref{sec:gen} and \ref{sec:spinvalley}. External factors that can alter the energetic landscape of such dipolar excitons, and consequently their photoluminescence or reflectance spectra, are beyond the scope of the present manuscript. Regarding the spin-valley physics and g-factors, previous studies focusing on intralayer excitons in monolayer TMDCs show that DFT calculations indeed provide a remarkable agreement to experimental data for pristine\cite{Robert2021PRL, Zinkiewicz2021NL} and strained\cite{FariaJunior2022NJP, Covre2022Nanoscale, Blundo2022PRL} cases. These intralayer excitons have larger conduction-to-valence band dipole matrix elements, larger electron-hole exchange contributions, and weaker dielectric screening when compared to interlayer excitons. Therefore, if DFT is indeed capturing the relevant spin-valley properties of intralayer excitons, it will be sufficient enough to provide a reliable description of the spin-valley properties in dipolar excitons. In fact, DFT studies on interlayer exciton g-factors\cite{Wozniak2020PRB, Forste2020NatComm, Xuan2020PRR} are indeed capable of providing reliable comparison with experiments.

%--------------------------------------------------------------------------------

\subsection{Symmetry-based selection rules}
\label{sec:inter_symmetry}

Starting from the symmetry perspective, we can use the irreps computed in Section~\ref{sec:gen} to characterize the optical selection rules of the possible interlayer excitons. Let us first discuss the selection rules of direct excitons at the K points, in order to compare with and validate our DFT calculations in the next Section~\ref{sec:inter_Efield}. Direct optical transitions are characterized by the matrix element mediated by the momentum operator
\begin{equation}
\left\langle i \left| \vec{p}\cdot \hat{\epsilon} \right| f\right\rangle \sim \mathbb{I}^*_i \otimes \mathbb{I}_p \otimes \mathbb{I}_f \, ,
\end{equation}
in which $\mathbb{I}_{i(f)}$ is the irrep of the initial (final) state and $\mathbb{I}_{p}$ is the irrep of the momentum operator for a specific polarization ($\hat{\epsilon} = \{ \sigma_+, \sigma_-, z \}$). The components of the momentum operator for circularly and linearly polarized light in the $C_{3}$ group transforms as 
\begin{equation}
K_{1}\sim z,\;K_{2}\sim\sigma_{-},\;K_{3}\sim\sigma_{+} \, .
\label{eq:irrep_p}
\end{equation}

Combining Eq.~\ref{eq:irrep_p} with the irreps of valence and conduction bands given in Table~\ref{tab:irreps}, we can compute the  optical selection rules for K-K dipolar excitons. The results are shown in Table~\ref{tab:interKK_selection_rules} and agree with previous symmetry-based and DFT calculations\cite{Yu2017SciAdv, Wu2018PRB, Wozniak2020PRB}. It is also possible to verify that the \textit{intralayer} optical selection rules are nicely preserved. For instance, transitions from $\textrm{v}_{\textrm{W}}(\textrm{K})$ to $\textrm{c}_{\textrm{W}+}(\textrm{K})$ provide $\sigma_+$, whereas transitions from $\textrm{v}_{\textrm{M}}(\textrm{K})$ to $\textrm{c}_{\textrm{M}-}(\textrm{K})$ provide $\sigma_+$ for R-type stackings and $\sigma_-$ for H-type stackings. One important point to discuss here is related to the so-called \textit{spin-dark} optical transitions (sometimes also refereed to as  \textit{spin-flip} or \textit{spin-singlet} transitions), for instance, as the $\textrm{v}_{\textrm{W}}(\textrm{K}) \rightarrow \textrm{c}_{\textrm{M}-}(\textrm{K})$ transition in the \Hh~stacking. The orientation of $S_z$, given in Tables~\ref{tab:SLmono} and \ref{tab:SLhet}, is not a indicative of allowed or forbidden optical transitions. In fact, $S_z \neq 1$ if we increase the number digits after the decimal point in Tables~\ref{tab:SLmono} and \ref{tab:SLhet}. This is already evident in the monolayer case due to the existence of the dark excitons\cite{Robert2017PRB, Molas2019PRL, Zinkiewicz2020Nanoscale}. The main point here is that SOC always introduces a mixing of spins (even in graphene\cite{Kurpas2019PRB}) and that the momentum matrix element does not flip spin (more discussions about spin-mixing and optical transitions for \textit{spin-dark} states in monolayer TMDCs can be found in Ref.\cite{FariaJunior2022NJP}). Furthermore, some of the interlayer optical transitions presented in Table~\ref{tab:interKK_selection_rules} have the same structure as the dark/grey excitons in monolayers\cite{Robert2017PRB, Molas2019PRL, Zinkiewicz2020Nanoscale}, i. e., one transition with $z$ polarization and the other with $\sigma_{\pm}$. This feature actually explains the energy splitting of $\sim$0.1 meV experimentally observed at zero magnetic field by Li and coauthors\cite{Li2022arXiv} in atomically reconstructed \MS/\WS~heterobilayers, akin to the dark-grey exciton splitting of $\sim$0.5 meV in pristine W-based monolayers\cite{Robert2017PRB, Molas2019PRL, Robert2017PRB}.

\begin{table}
\caption{Symmetry-based selection rules for the direct interlayer excitons at the K point in the $C_3$ symmetry group.}
\begin{center}
\begin{tabular}{lll}
\hline
\hline
 & $\textrm{v}_{\textrm{W}}(\textrm{K}) \rightarrow \textrm{c}_{\textrm{M}-}(\textrm{K})$ & $\textrm{v}_{\textrm{W}}(\textrm{K}) \rightarrow \textrm{c}_{\textrm{M}+}(\textrm{K})$ \\ [1mm]
\hline
\Rh & $K_{4}^{*}\otimes K_{5}=K_{3}\Rightarrow\sigma_{+}$ & $K_{4}^{*}\otimes K_{4}=K_{1}\Rightarrow z$\\
\RM & $K_{5}^{*}\otimes K_{5}=K_{1}\Rightarrow z$ & $K_{5}^{*}\otimes K_{4}=K_{2}\Rightarrow\sigma_{-}$\\
\RX & $K_{6}^{*}\otimes K_{5}=K_{2}\Rightarrow\sigma_{-}$ & $K_{6}^{*}\otimes K_{4}=K_{3}\Rightarrow\sigma_{+}$ \\ [1mm]
\hline
\Hh & $K_{6}^{*}\otimes K_{4}=K_{3}\Rightarrow\sigma_{+}$ & $K_{6}^{*}\otimes K_{5}=K_{2}\Rightarrow\sigma_{-}$\\
\HM & $K_{5}^{*}\otimes K_{4}=K_{2}\Rightarrow\sigma_{-}$ & $K_{5}^{*}\otimes K_{5}=K_{1}\Rightarrow z$\\
\HX & $K_{4}^{*}\otimes K_{4}=K_{1}\Rightarrow z$ & $K_{4}^{*}\otimes K_{5}=K_{3}\Rightarrow\sigma_{+}$\\
\hline
\hline
\end{tabular}
\end{center}
\label{tab:interKK_selection_rules}
\end{table}

For optical transitions involving momentum-indirect dipolar excitons, i. e., transitions involving valence and conduction bands at different $\vec{k}$ points, a phonon is required to ensure momentum conservation. The optical transition matrix element mediated by phonons is written as\cite{inui}
\begin{equation}
\sum_{l}\left\langle i\left|H_{\text{e-ph}}\right|l\right\rangle \left(E_{f}-E_{l}\right)^{-1}\left\langle l\left|  \vec{p}\cdot \hat{\epsilon} \right|f\right\rangle \sim \mathbb{I}_{i}^{*} \otimes \mathbb{I}_{\text{ph}}\otimes \mathbb{I}_{p} \otimes \mathbb{I}_{f} \, ,
\end{equation}
in which $H_{\text{e-ph}}$ takes into account the electron-phonon coupling and transforms as the irrep $I_{\text{ph}}$. It is beyond the scope of this manuscript to provide deeper insights into the strength of the electron-phonon coupling but rather discuss qualitatively the possible symmetry-allowed mechanisms involving the momentum-indirect dipolar excitons.

\begin{table}
\caption{Direct product of valence bands at the $-$K point and conduction bands at the K point in the $C_3$ symmetry group.}
\begin{center}
\begin{tabular}{lcc}
\hline
\hline
 & $\textrm{v}_{\textrm{W}}(-\textrm{K})\rightarrow\textrm{c}_{\textrm{M}-}(\textrm{K})$ & $\textrm{v}_{\textrm{W}}(-\textrm{K})\rightarrow\textrm{c}_{\textrm{M}+}(\textrm{K})$ \\ [1mm]
 \hline
\Rh & $K_{5}^{*}\otimes K_{5}=K_{1}$ & $K_{5}^{*}\otimes K_{4}=K_{2}$\\
\RM & $K_{4}^{*}\otimes K_{5}=K_{3}$ & $K_{4}^{*}\otimes K_{4}=K_{1}$\\
\RX & $K_{6}^{*}\otimes K_{5}=K_{2}$ & $K_{6}^{*}\otimes K_{4}=K_{3}$  \\ [1mm]
\hline
\Hh & $K_{6}^{*}\otimes K_{4}=K_{3}$ & $K_{6}^{*}\otimes K_{5}=K_{2}$\\
\HM & $K_{4}^{*}\otimes K_{4}=K_{1}$ & $K_{4}^{*}\otimes K_{5}=K_{3}$\\
\HX & $K_{5}^{*}\otimes K_{4}=K_{2}$ & $K_{5}^{*}\otimes K_{5}=K_{1}$\\
\hline
\hline
\end{tabular}
\end{center}
\label{tab:interKpKm_selection_rules}
\end{table}

For indirect excitons with electrons and holes in opposite K valleys (-K and K, sharing the same $C_3$ group), the direct product $\mathbb{I}_{i}^{*} \otimes \mathbb{I}_{f}$ provides the output shown in Table~\ref{tab:interKpKm_selection_rules}. Notice that the output is always restricted to the irreps $K_1$, $K_2$, and $K_3$. Phonons in the $C_3$ group would span all the irreps of the simple group, namely $K_1$, $K_2$, and $K_3$. Combining the results given in Table~\ref{tab:interKpKm_selection_rules} and performing a direct product with all possible phonon modes, we obtain
\begin{align}
K_{1}\otimes\left\{ K_{1},K_{2},K_{3}\right\}  & =\left\{ K_{1},K_{2},K_{3}\right\} \nonumber \\
K_{2}\otimes\left\{ K_{1},K_{2},K_{3}\right\}  & =\left\{ K_{2},K_{3},K_{1}\right\} \nonumber \\
K_{3}\otimes\left\{ K_{1},K_{2},K_{3}\right\}  & =\left\{ K_{3},K_{1},K_{2}\right\} 
\end{align}
which always provides $K_1$, $K_2$, and $K_3$. These irreps also encode the light polarization, as given in Eq.~\ref{eq:irrep_p}. What our results suggest is that any type of phonon with momentum K can, in principle, mediate the optical recombination of dipolar excitons with electrons and holes residing in opposite K valleys. Perhaps some of these processes are not favored due to off-resonant conditions and weak electron-phonon coupling, but could be effectively triggered once phonon energies are brought into resonance, such as the case of the 24T anomaly\cite{Nagler2017NatComm, Delhomme2020TDM, Smirnov2022TDM}. Certainly, these phonon processes are intimately dependent on the strength of the electron-phonon couplings and how those couplings are translated for the final excitonic state (based on the particular conduction and valence bands constituents).

\begin{table}
\caption{Direct product of valence bands at the $\Gamma$ point and conduction bands at the K point in the $C_3$ symmetry group.}
\begin{center}
\begin{tabular}{lcc}
\hline
\hline
 & \Rh,\RM,\RX & \Hh,\HM,\HX \\ [1mm]
 \hline
$\textrm{v}_{\Gamma}(K_{4})\rightarrow\textrm{c}_{\textrm{M}-}(\textrm{K})$ & $K_{4}^{*}\otimes K_{5}=K_{3}$ & $K_{4}^{*}\otimes K_{4}=K_{1}$\\
$\textrm{v}_{\Gamma}(K_{5})\rightarrow\textrm{c}_{\textrm{M}-}(\textrm{K})$ & $K_{5}^{*}\otimes K_{5}=K_{1}$ & $K_{5}^{*}\otimes K_{4}=K_{2}$ \\ [1mm]
\hline 
$\textrm{v}_{\Gamma}(K_{4})\rightarrow\textrm{c}_{\textrm{M}+}(\textrm{K})$ & $K_{4}^{*}\otimes K_{4}=K_{1}$ & $K_{4}^{*}\otimes K_{5}=K_{3}$\\
$\textrm{v}_{\Gamma}(K_{5})\rightarrow\textrm{c}_{\textrm{M}+}(\textrm{K})$ & $K_{5}^{*}\otimes K_{4}=K_{2}$ & $K_{5}^{*}\otimes K_{5}=K_{1}$\\
\hline
\hline
\end{tabular}
\end{center}
\label{tab:interG_selection_rules}
\end{table}

\begin{table*}
\caption{Dipolar exciton g-factors involving the top valence band states (at $\Gamma$, K, and $-$K points) to conduction bands of Mo at the K point ($\textrm{c}_{\textrm{M}\pm}$), or to the conduction bands at the Q point ($\textrm{c}_{\textrm{Q}\pm}$) at zero electric field and equilibrium interlayer distance. The g-factors with unambiguously determined signs are given in bold.}
\begin{center}
\begin{tabular}{rcccccc}
\hline
\hline
 & \Rh & \RM & \RX & \Hh & \HM & \HX \\ [1mm]
 \hline
$\ensuremath{\textrm{v}_{\textrm{W}}(\textrm{K})\rightarrow\textrm{c}_{\textrm{M}-}(\textrm{K})}$ & $\mathbf{-6.42}$ & $6.34$ & $\mathbf{+6.34}$ & $\mathbf{-17.18}$ & $\mathbf{+17.62}$ & $17.66$\\
$\ensuremath{\textrm{v}_{\textrm{W}}(\textrm{K})\rightarrow\textrm{c}_{\textrm{M}+}(\textrm{K})}$ & $10.93$ & $\mathbf{+10.85}$ & $\mathbf{-10.84}$ & $\mathbf{+12.70}$ & $13.10$ & $\mathbf{-13.18}$\\
$\ensuremath{\textrm{v}_{\textrm{W}}(\textrm{K})\rightarrow\textrm{c}_{\textrm{Q}-}(\textrm{Q})}$ & $10.01$ & $9.78$ & $9.94$ & $9.56$ & $10.49$ & $10.22$\\
$\ensuremath{\textrm{v}_{\textrm{W}}(\textrm{K})\rightarrow\textrm{c}_{\textrm{Q}+}(\textrm{Q})}$ & $13.67$ & $13.34$ & $13.42$ & $12.74$ & $12.94$ & $12.94$  \\ [1mm]
\hline
$\ensuremath{\textrm{v}_{\textrm{W}}(-\textrm{K})\rightarrow\textrm{c}_{\textrm{M}-}(\textrm{K})}$ & $17.64$ & $17.63$ & $17.57$ & $5.99$ & $6.43$ & $6.41$\\
$\ensuremath{\textrm{v}_{\textrm{W}}(-\textrm{K})\rightarrow\textrm{c}_{\textrm{M}+}(\textrm{K})}$ & $13.13$ & $13.12$ & $13.07$ & $10.47$ & $10.94$ & $10.89$\\
$\ensuremath{\textrm{v}_{\textrm{W}}(-\textrm{K})\rightarrow\textrm{c}_{\textrm{Q}-}(\textrm{Q})}$ & $14.05$ & $14.19$ & $13.97$ & $13.61$ & $13.55$ & $13.85$\\
$\ensuremath{\textrm{v}_{\textrm{W}}(-\textrm{K})\rightarrow\textrm{c}_{\textrm{Q}+}(\textrm{Q})}$ & $10.39$ & $10.63$ & $10.49$ & $10.43$ & $11.10$ & $11.12$ \\ [1mm]
\hline
$\ensuremath{\textrm{v}_{\Gamma_{1}}\rightarrow\textrm{c}_{\textrm{M}-}(\textrm{K})}$ & $3.58$ & $3.67$ & $3.62$ & $7.58$ & $7.62$ & $7.62$\\
$\ensuremath{\textrm{v}_{\Gamma_{1}}\rightarrow\textrm{c}_{\textrm{M}+}(\textrm{K})}$ & $0.92$ & $0.85$ & $0.87$ & $3.10$ & $3.10$ & $3.14$\\
$\ensuremath{\textrm{v}_{\Gamma_{1}}\rightarrow\textrm{c}_{\textrm{Q}-}(\textrm{Q})}$ & $0.01$ & $0.23$ & $0.03$ & $0.03$ & $0.49$ & $0.18$\\
$\ensuremath{\textrm{v}_{\Gamma_{1}}\rightarrow\textrm{c}_{\textrm{Q}+}(\textrm{Q})}$ & $3.66$ & $3.33$ & $3.46$ & $3.14$ & $2.94$ & $2.90$ \\ [1mm]
\hline
$\ensuremath{\textrm{v}_{\Gamma_{2}}\rightarrow\textrm{c}_{\textrm{M}-}(\textrm{K})}$ & $7.63$ & $7.63$ & $7.60$ & $3.60$ & $3.58$ & $3.63$\\
$\ensuremath{\textrm{v}_{\Gamma_{2}}\rightarrow\textrm{c}_{\textrm{M}+}(\textrm{K})}$ & $3.13$ & $3.12$ & $3.11$ & $0.88$ & $0.94$ & $0.85$\\
$\ensuremath{\textrm{v}_{\Gamma_{2}}\rightarrow\textrm{c}_{\textrm{Q}-}(\textrm{Q})}$ & $4.04$ & $4.19$ & $4.01$ & $4.01$ & $3.55$ & $3.81$\\
$\ensuremath{\textrm{v}_{\Gamma_{2}}\rightarrow\textrm{c}_{\textrm{Q}+}(\textrm{Q})}$ & $0.39$ & $0.63$ & $0.52$ & $0.83$ & $1.10$ & $1.08$\\
\hline
\hline
\end{tabular}
\end{center}
\label{tab:inter_gfactors}
\end{table*}

To investigate indirect dipolar excitons, with holes at the $\Gamma$ point and electrons at the K valley which do not have the same symmetry group, we must describe both conduction and valence band states in the same footing using compatibility relations\cite{dresselhaus}. It is more convenient to go from a higher symmetry group to a lower symmetry group then the other way around. With this in mind, we mapped $\Gamma$ point states to the K valley $\Gamma_4 \rightarrow K_4 \oplus K_5$ and present the direct product in Table~\ref{tab:interG_selection_rules}. The same argument discussed in the case of electrons and holes at opposite valleys apply here, and multiple phonons with momentum K could trigger the optical recombination of such $\Gamma$-K excitons.

Finally, we discuss the indirect excitons involving conduction bands at the Q point and holes at either $\Gamma$ or K points. Since the $C_1$ symmetry group contains a single irrep in the simple group, it means that all electronic states, phonon modes and operators are mapped into this irrep. Consequently, multiple phonon mediated processes are allowed. We note that Q-K transitions have been recently observed\cite{Barre2022Science}, supporting the idea that some of these symmetry-allowed phonons discussed here are indeed favoring the optical recombination.

%--------------------------------------------------------------------------------

\subsection{Effective g-factors}
\label{sec:inter_gfactors}

Combining the band g-factors studied in Section~\ref{sec:spinvalley} with the optical selection rules discussed in Section~\ref{sec:inter_symmetry}, we investigate here the effective g-factors of the relevant dipolar excitons. Since the momentum-indirect excitons require a phonon to mediate the optical emission, we can only determine unambiguously the sign of the g-factors for direct interlayer excitons, while for the other dipolar states we restrict ourselves to the magnitude of the g-factors. The dipolar exciton g-factors can be written as
\begin{equation}
g(X) = 2 \left| g(c) \pm g(v) \right|
\label{eq:gX}
\end{equation}
in which $g(c)$ and $g(v)$ represent the conduction and valence band g-factors, and the $\pm$ sign contemplates the two possibilities of time-reversal partners. For example, interlayer excitons with the valence band $v=\textrm{v}_\textrm{W}$ and the conduction bands $c=\textrm{c}_{\textrm{Mo}\pm}$ could be generated with $v$ and $c$ at the K point (direct in momentum), as indicated in Fig.\ref{fig:bs_qtl_GKM_scheme}g or, with $v$ in the K point and $c$ in the $-$K point (indirect in momentum).

\begin{figure*}
\begin{center}
\includegraphics{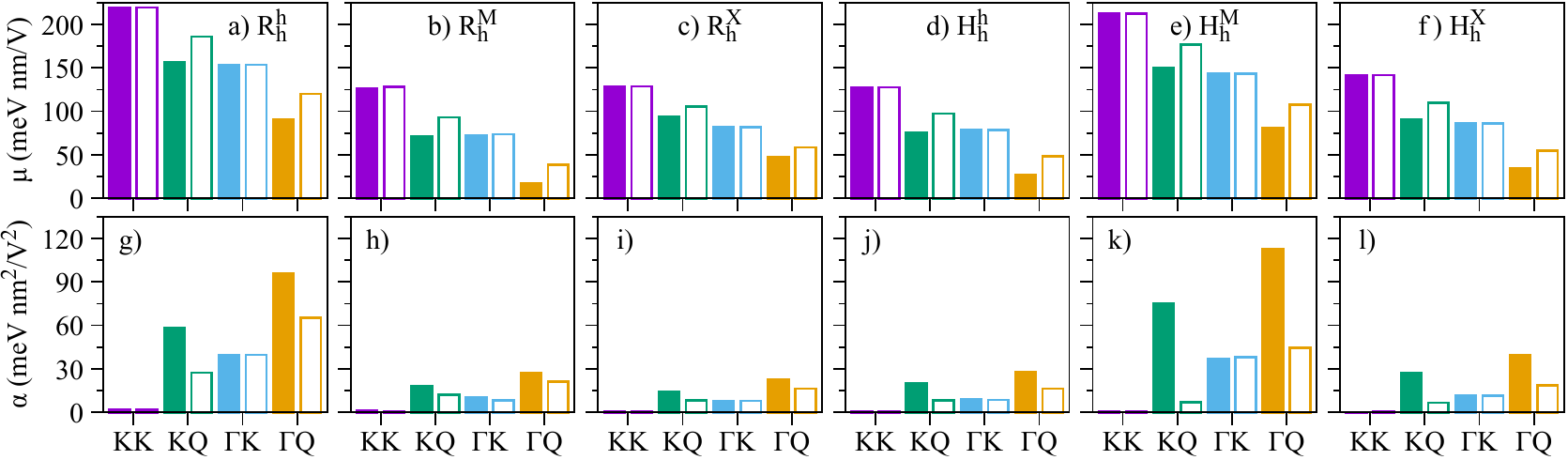}
\end{center}
\caption{Calculated values of the electric dipole moments for a) \Rh, b) \RM, c) \RX, d) \Hh, e) \HM~and f) \HX~stackings. Calculated values of the polarizabilities for g) \Rh, h) \RM, i) \RX, j) \Hh, k) \HM~and l) \HX~stackings. The x-axis indicate the type of dipolar excitons (see Fig.~\ref{fig:bs_qtl_GKM_scheme}g. The values originating from $\textrm{c}_{\textrm{M/Q}-}$ ($\textrm{c}_{\textrm{M/Q}+}$) are shown with colored (open) boxes.}
\label{fig:Efield_interlayer_mu_alpha}
\end{figure*}

The calculated g-factor values for the studied dipolar excitons are shown in Table~\ref{tab:inter_gfactors}. The largest absolute values stem from the KK dipolar excitons ($v$ at K and $c$ at K in H-type, $v$ at -K and $c$ at K in R-type) since they involve the energy bands with largest g-factors. Some dipolar excitons show very similar g-factors, such as the values related to $\ensuremath{\textrm{v}_{\textrm{W}}(-\textrm{K})\rightarrow\textrm{c}_{\textrm{M}+}(\textrm{K})}$ and $\ensuremath{\textrm{v}_{\textrm{W}}(-\textrm{K})\rightarrow\textrm{c}_{\textrm{M}+}(\textrm{K})}$ excitons in H-type stackings. We also observed g-factors with values $<1$ and even $\sim0$ [$\ensuremath{\textrm{v}_{\Gamma_{1}}\rightarrow\textrm{c}_{\textrm{Q}-}(\textrm{Q})}$], mainly originating from the valence bands at the $\Gamma$ point These different g-factors originated from different stackings and distinct dipolar exciton species could be observed in the cross-over region between deep and shallow moiré potentials as a function of the twist angle or temperature, however, no systematic experimental study has been realized to explore such conditions. We emphasize that our calculated values are in excellent agreement with previously reported theoretical studies\cite{Wozniak2020PRB, Xuan2020PRR, Gillen2021pssb, Zhao2022}, and lie within the available experimental values, which range from -15 to -17 in H-type stackings\cite{Nagler2017NatComm, Seyler2019Nature, Wang2020NL, Delhomme2020TDM, Mahdikhanysarvejahany2021npj, Forg2021NatComm, Holler2022PRB, Smirnov2022TDM, Li2022arXiv}, and from 5 to 8 in R-type stackings\cite{Seyler2019Nature, Ciarrocchi2019NatPhot, Joe2021PRB, Mahdikhanysarvejahany2021npj, Holler2022PRB, Smirnov2022TDM, Li2022arXiv, Qian2023}.

%--------------------------------------------------------------------------------

\subsection{Electric field dependence}
\label{sec:inter_Efield}

We now turn to the behavior of the different dipolar excitons as a function of the electric field. Motivated by the experimentally accessible regime\cite{Ciarrocchi2019NatPhot, Baek2020SciAdv, Shanks2021NL, Barre2022Science, Shanks2022NL}, in which there is no indication of any crossings in the top valence band or lower conduction bands, we restrict ourselves to electric field values $\gtrsim -0.75$ V/nm, ensuring that the conduction band of \MS~is always below the conduction band of \WS. For positive values of electric field, there are no visible band crossings, so we can extend our analysis towards electric field values of $\sim 2$ V/nm. In fact, some experimental studies that investigate dipolar excitons under electric field also show an asymmetric range\cite{Ciarrocchi2019NatPhot, Baek2020SciAdv, Barre2022Science}, supporting the electric field values we consider here.

The energy dependence of the dipolar excitons under external electric fields can be directly inferred from the dependence of the energy bands, because the effective mass are barely affected by the electric field (changes on the order of $10^{-3} - 10^{-2}$) and the dielectric environment remains unchanged. Therefore, the energy variation of the dipolar excitons with respect to the electric field can be extracted directly from the DFT calculations\cite{Plankl2021NatPhot} and modelled by the following dependence\cite{Roch2018NL, Chakraborty2019OME}
\begin{equation}
\Delta E (F_z) = -\mu F_z - \frac{\alpha}{2} F_z^2
\end{equation}
in which $\mu$ is the electric dipole moment (in meV nm/V), $\alpha$ is the polarizability (in meV nm$^2$/V$^2$), and $F_z$ is the electric field in V/nm.

\begin{figure*}
\begin{center}
\includegraphics{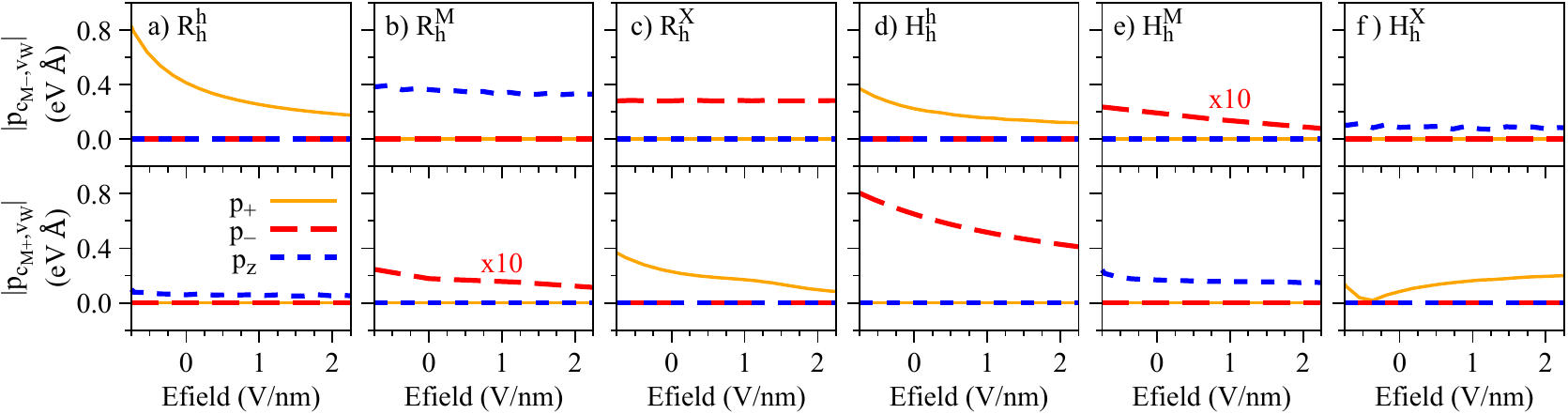}
\end{center}
\caption{Absolute value of the dipole matrix element for interlayer transitions at the K-point as a function of the electric field for a) \Rh, b) \RM, c) \RX, d) \Hh, e) \HM~and f) \HX~stackings.}
\label{fig:Efield_interlayer_dipole}
\end{figure*}

\begin{figure*}
\begin{center}
\includegraphics{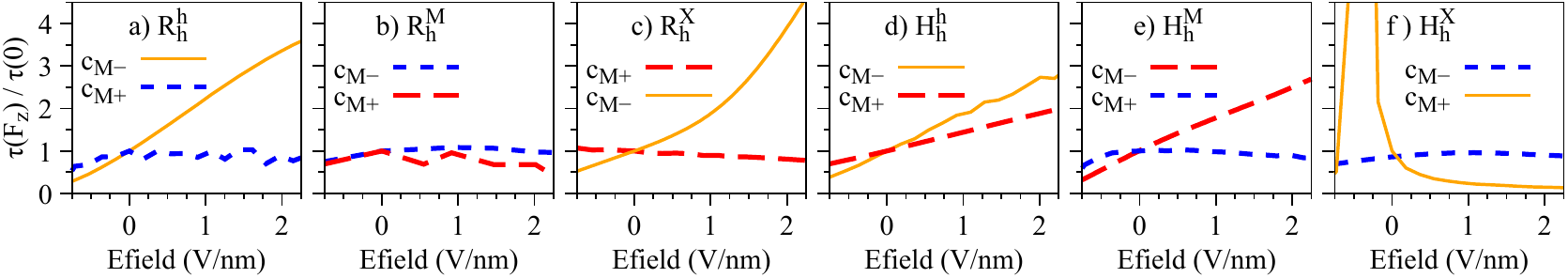}
\end{center}
\caption{Calculated ratio $\tau_{\text{rad}}(F_{z})/\tau_{\text{rad}}(0)$ as a function of the electric field for a) \Rh, b) \RM, c) \RX, d) \Hh, e) \HM~and f) \HX~stackings. We used $E_0 = 1.35$ eV for all cases and the calculated values of $\mu$ given in Fig.~\ref{fig:Efield_interlayer_mu_alpha}. The contribution of $\alpha$ is neglected since they are nearly vanishing for KK dipolar excitons.}
\label{fig:Efield_interlayer_radtime}
\end{figure*}

Our calculated values of the electric dipole moment and polarizabilities, fitted in the range of $-$0.7 to 0.7 V/nm, are presented in Fig.~\ref{fig:Efield_interlayer_mu_alpha}. Overall, our results reveal that the values of $\mu$ ($\alpha$) decrease (increase) as we go from momentum-direct to momentum-indirect dipolar excitons (KK $\rightarrow$ KQ $\rightarrow$ $\Gamma$K $\rightarrow$ $\Gamma$Q), reflecting the degree of layer delocalization of the bands involved ($\Gamma$ $>$ Q $>$ K, as shown in Fig.~\ref{fig:bs_qtl_GKM_scheme}a-f). These delocalization features have already been observed in the magnitude of interlayer excitons of \WS~homobilayer structures\cite{Altaiary2022NL, Huang2022PRB} and there is one report on \MS/\WS~heterobilayers\cite{Barre2022Science}. Our results also show that the spin-split conduction bands $\textrm{c}_{\textrm{M}\pm}$ have negligible effect on $\mu$ and $\alpha$ of KK and $\Gamma$K excitons, whereas the conduction bands $\textrm{c}_{\textrm{Q}\pm}$ provide different signatures since they have different delocalization over the two layers. Particularly for the KK dipolar excitons, we note that the calculated values of $\alpha$ are quite small ($0.2 - 2$ meV nm$^2$/V$^2$) and the two possible excitons, stemming from $\textrm{c}_{\textrm{M}\pm}$, are nearly indistinguishable under the electric field dependence. However, they would have different emission energies (due to the SOC splitting of $\textrm{c}_{\textrm{M}\pm}$) and distinct g-factor signatures. We note that our calculated values of $\mu$ for KK and KQ dipolar excitons, and particularly that $\mu_{\textrm{KK}} > \mu_{\textrm{KQ}}$ are consistent with recent GW + Bethe-Salpeter calculations and experiments in \MS/\WS~heterobilayers\cite{Barre2022Science}.

To validate the symmetry-based selection rules of direct interlayer excitons discussed in Section~\ref{sec:inter_symmetry}, we present the dipole matrix elements computed within DFT in Fig.~\ref{fig:Efield_interlayer_dipole}. All the nonzero values match exactly the symmetry-based arguments discussed in Section~\ref{sec:inter_iX}. Generally, the dipole matrix elements decrease for increasing values of the electric field, with the exception of the $\textrm{p}_{\textrm{c}_{\textrm{M}+},\textrm{v}_{\textrm{W}}}$ that slightly increases for increasing electric field, reflecting the impact of the atomic registry on the wave functions of van der Waals heterostacks. Furthermore, the \textit{spin-dark} optical transitions discussed in terms of symmetry in Section~\ref{sec:inter_symmetry} are also present here. The momentum operator that enters the dipole matrix element (Eq.~\ref{eq:irrep_p}) does not flip spin and therefore this transition is optically bright only if the 2-component spinors are mixed, which is indeed the case in our first-principles calculations with SOC. Our calculated values of the dipole matrix elements, along with their respective polarization, can be readily used to generate massive Dirac models for the dipolar excitons via the first order perturbation within the \textbf{k.p} formalism, allowing the investigation of spin-dependent physics beyond the parabolic approximation\cite{Wu2018PRB, Erkensten2021PRB}.

Based on the calculated values of the energy dependence and dipole matrix elements with respect to the electric field, we can evaluate the radiative lifetime for direct dipolar excitons, given by\cite{Glazov2014PRB, Glazov2015pssb}
\begin{equation}
\tau_{\text{rad}}=\frac{\sqrt{\varkappa_{b}}c\hbar^{2}}{2\pi e^{2}}\frac{E_{0}}{\left|p_{c,v}\right|^{2}\left|\phi(0)\right|^{2}} \, ,
\end{equation}
with $\varkappa_{b}$ being high-frequency dielectric constant of the environment, $c$ is the speed of light, $\hbar$ is the reduced Plank constant, $e$ is the electron charge, $E_0$ is the exciton energy, and $\left|\phi(0)\right|^{2}$ is the exciton envelope function evaluated at the origin. The relevant quantities that depend on the electric field, $F_z$, are the exciton energy, $E_{0}\rightarrow E_{0}+\Delta E(F_{z})$ (see Eq.~\ref{eq:EZ}) and the dipole matrix element, $p_{c,v}\rightarrow p_{c,v}(F_{z})$ (see Fig.~\ref{fig:Efield_interlayer_dipole}). Typical interlayer excitons exhibit lifetimes reaching almost 100 ns at very low temperatures and a few ns at room temperature\cite{Rivera2015NatComm, Miller2017NL, Nagler2017TDM}, with roughly one order of magnitude fluctuation when comparing different samples. To understand the isolated impact of the electric field on the radiative lifetime of the direct dipolar excitons, we consider the quantity

\begin{equation}
\frac{\tau_{\text{rad}}(F_{z})}{\tau_{\text{rad}}(0)}=\frac{E_{0}+\Delta E(F_{z})}{\left|p_{c,v}(F_{z})\right|^{2}}\frac{\left|p_{c,v}(0)\right|^{2}}{E_{0}} \, ,
\end{equation}
and display its calculated values in Fig.~\ref{fig:Efield_interlayer_radtime} as function of the electric field. The calculated ratio $\tau_{\text{rad}}(F_{z})/\tau_{\text{rad}}(0)$ varies in general in the range of 0.5 to 4.5, except for the case \HX~that shows a drastic increase of the radiative lifetime because the dipole matrix element decreases to nearly zero in Fig.~\ref{fig:Efield_interlayer_dipole}f. Particularly for the high-symmetry stacking \Hh, which is the most favorable one for $\sim$60\degree~samples, both direct dipolar excitons (singlet and triplet) exhibit an increase in the radiative lifetime as the electric field increases, which can be accessed experimentally in current devices.

\begin{figure*}
\begin{center}
\includegraphics{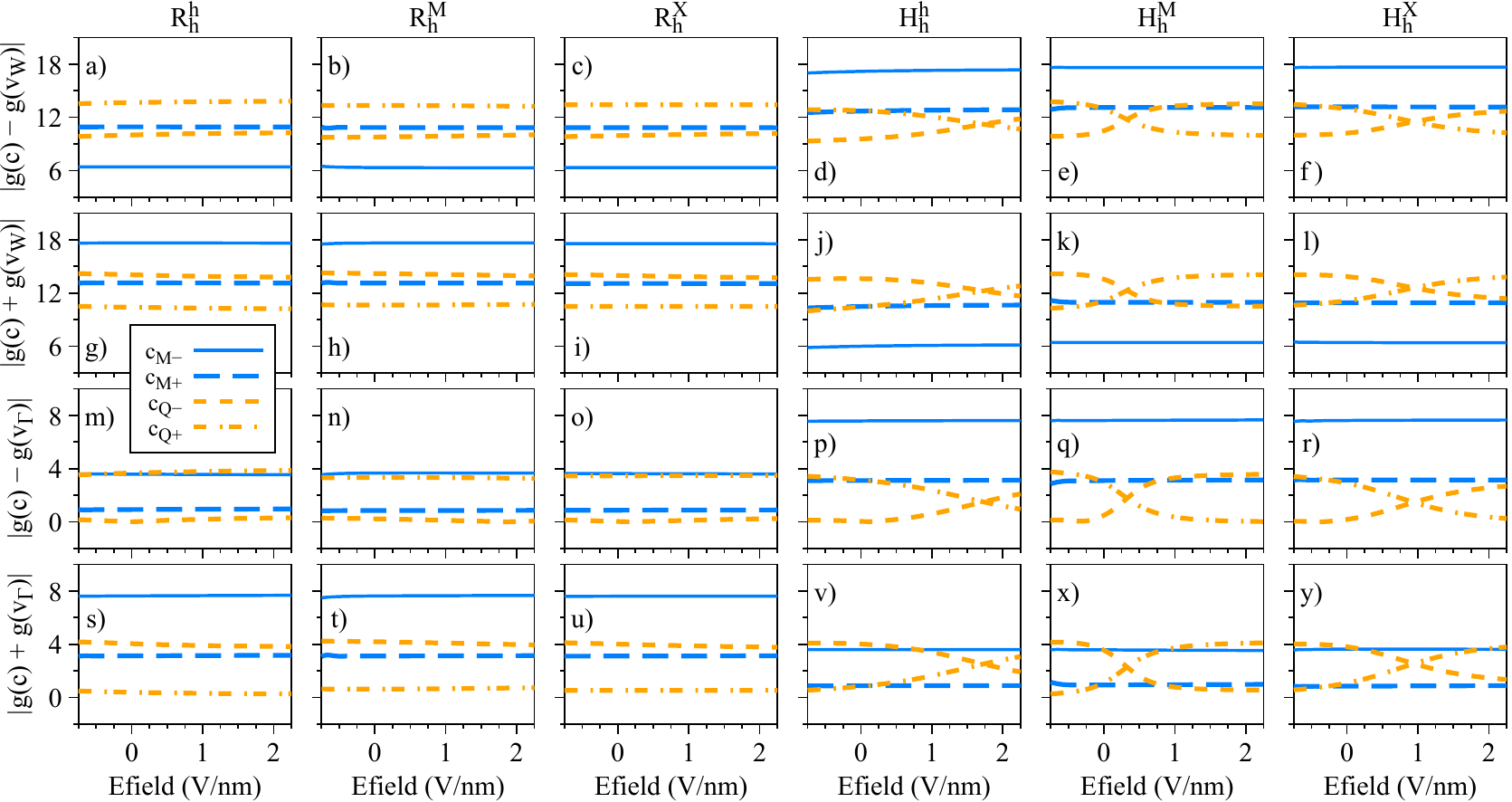}
\end{center}
\caption{Dipolar exciton g-factors as a function of the electric field for the R- and H-type systems studied for the cases a-f) $\left|g(c)-g(\textrm{v}_\textrm{W})\right|$, g-l) $\left|g(c)+g(\textrm{v}_\textrm{W})\right|$, m-r) $\left|g(c)-g(\textrm{v}_\Gamma)\right|$ and s-y) $\left|g(c)+g(\textrm{v}_\Gamma)\right|$.}
\label{fig:Efield_interlayer_gfactors_RandH}
\end{figure*}

Let us now discuss electric field effects on the valley Zeeman physics of the dipolar excitons (schematically shown in Fig.~\ref{fig:bs_qtl_GKM_scheme}g). The effective g-factors calculated via Eq.~\ref{eq:gX} as a function of the electric field are shown in Fig.~\ref{fig:Efield_interlayer_gfactors_RandH} for all the considered R- and H-type stackings. Our results reveal that these dipolar excitons have a very robust valley Zeeman, barely changing under external electric field. The only exception are the dipolar excitons involving Q bands in the H-type stackings. Because of the g-factor crossing between $\textrm{c}_{\textrm{Q}+}$ and $\textrm{c}_{\textrm{Q}-}$ bands (see Figs.~\ref{fig:Efield_gz}p,q,r) stemming from spin mixing effects (see Figs.~\ref{fig:Efield_Sz}p,q,r and Figs.~\ref{fig:Efield_Lz}p,q,r), the valley Zeeman of these indirect dipolar excitons is consequently affected and exhibits pronounced changes for increasing electric fields (see Figs.~\ref{fig:Efield_interlayer_gfactors_RandH}d-f,j-l,p-r,v-y). Therefore, without significant band hybridization the dipolar excitons retain their valley Zeeman signatures over a large range of applied electric field values, which can be particularly relevant for electrically-driven opto-spintronics applications\cite{Sierra2021NatNano}. In fact, our results reveal very fundamental and general features about the valley Zeeman mixing of excitonic states (including dipolar excitons) in 2D materials and van der Waals heterostructures, in this case, stemming from the band hybridization between different layers. Clear signatures of the valley Zeeman mixing on the excitonic states are still rather limited, but have been already observed experimentally in a few systems, namely, strained WS$_2$ monolayers\cite{Blundo2022PRL}, defect-mediated excitons in MoS$_2$ monolayers\cite{Amit2022,Hotger2022}, WS$_2$/graphene heterostructures\cite{FariaJunior2023} and electrically-controlled trilayer MoSe$_2$ structures\cite{Feng2022}.

%--------------------------------------------------------------------------------

\subsection{Interlayer distance variation}
\label{sec:inter_iX}

\begin{figure*}
\begin{center}
\includegraphics{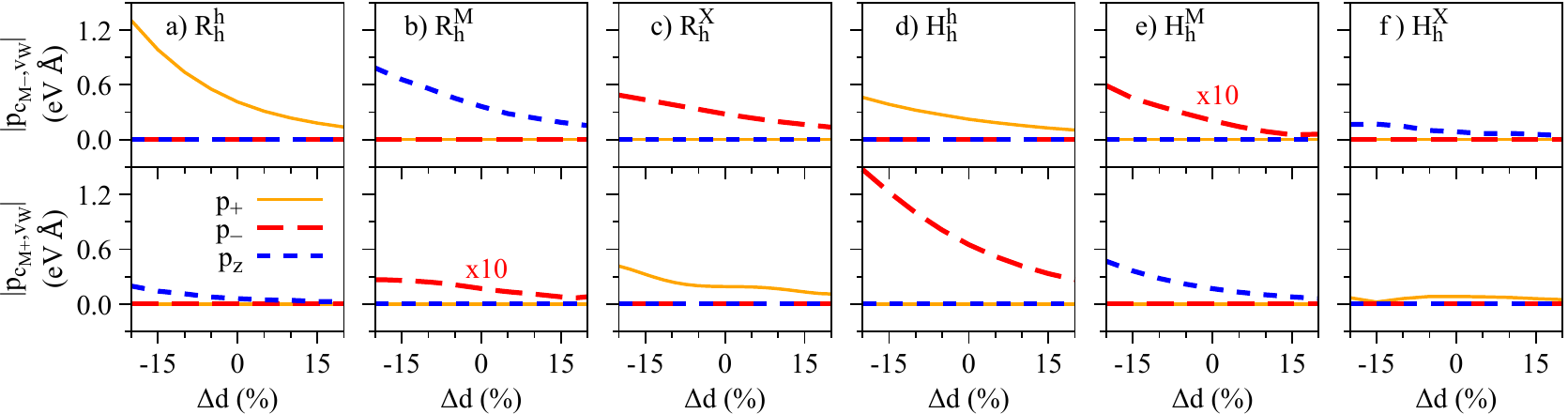}
\end{center}
\caption{Same as Fig.~\ref{fig:Efield_interlayer_dipole} but as a function o the interlayer distance.}
\label{fig:iX_interlayer_dipole}
\end{figure*}

\begin{figure*}
\begin{center}
\includegraphics{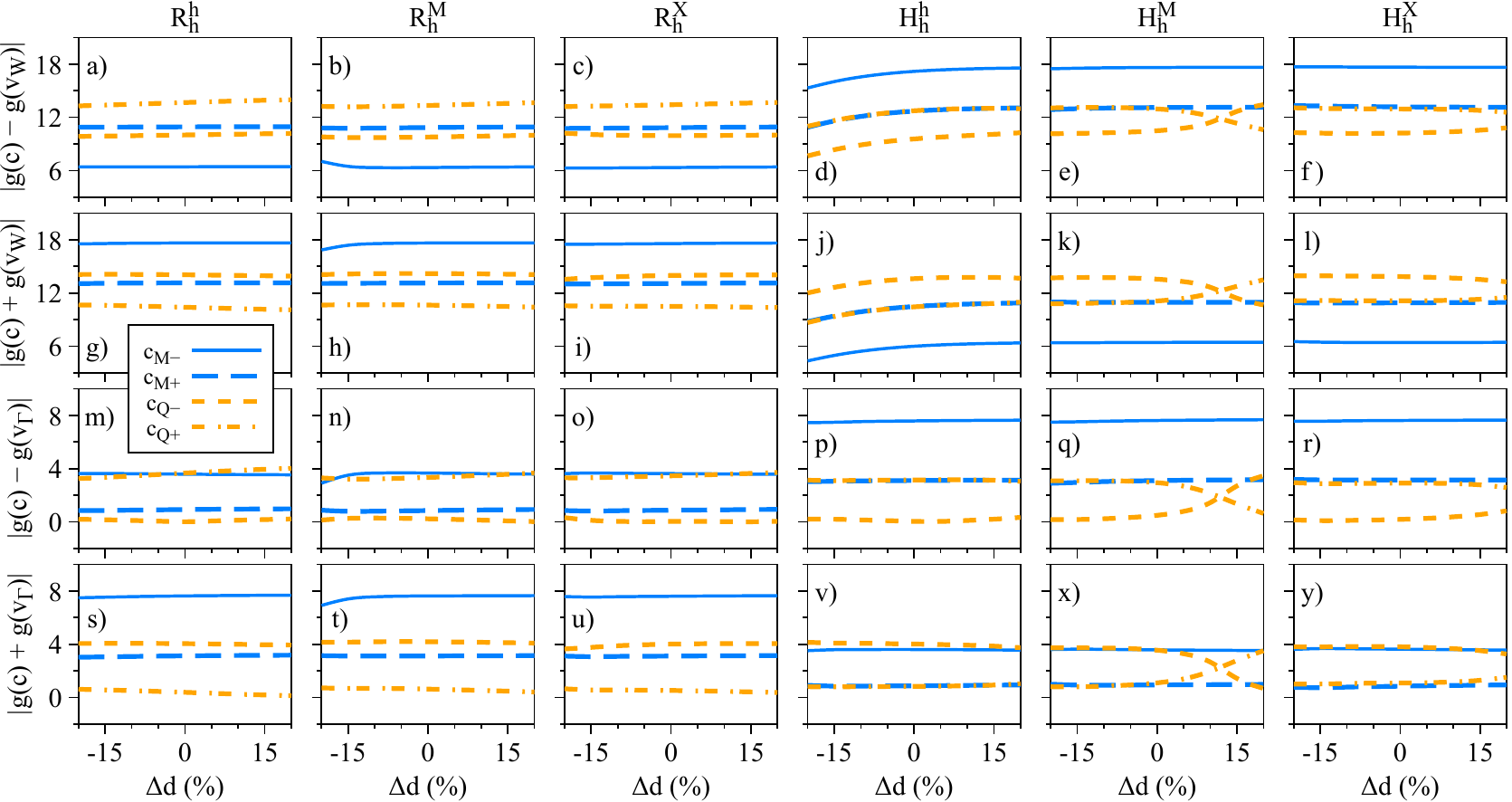}
\end{center}
\caption{Same as Fig.~\ref{fig:Efield_interlayer_gfactors_RandH} but as a function of the interlayer distance.}
\label{fig:iX_interlayer_gfactors_RandH}
\end{figure*}

In this section we focus on the effect of the interlayer distance variation. Our calculated values of the dipole matrix elements and the g-factors are presented in Fig.~\ref{fig:iX_interlayer_dipole} and Fig.~\ref{fig:iX_interlayer_gfactors_RandH}, respectively, and follow the same convention as discussed in the case of the electric field (Figs.~\ref{fig:Efield_interlayer_dipole} and \ref{fig:Efield_interlayer_gfactors_RandH}). Particularly, we found a strong decrease of the dipole matrix elements as the interlayer distance increases, consequently making the dipolar excitons less optically active. These findings provide support to the idea that weakly coupled heterobilayers (with an interlayer separation larger than the equilibrium distance), with perhaps additional electric field displacements due to asymmetric bottom and top environments, may exhibit highly quenched optical emission of dipolar excitons\cite{Plankl2021NatPhot, Parzefall2021TDM}. Furthermore, since the dipole matrix elements behave in a similar fashion as the electric field we expect similar behavior for the quantity $\tau_{\text{rad}}(F_{z})/\tau_{\text{rad}}(0)$. Another interesting aspect, visible in the \Hh~stacking, is the decrease of the valley Zeeman originating from the top valence band at the K point, $\textrm{v}_\textrm{W}$, as the interlayer distance decreases (Figs.~\ref{fig:iX_interlayer_gfactors_RandH}d,j). We believe this feature can be experimentally probed using external pressure\cite{Fulop2021npj2D, Oliva2022ACSami}, which effectively decreases the interlayer distance. For the \HM~stacking, we also observe the signatures of Q-band mixing and reversal of the g-factor values as the interlayer distance increases.

%================================================================================

\section{Concluding remarks}
\label{sec:conclusions}

In summary, we performed detailed first-principles calculations on the spin-valley physics of \MS/\WS~heterobilayers under the effect of out-of-plane electric fields and as a function of the interlayer separation. Specifically, we investigated the spin and orbital degrees of freedom, g-factors, and symmetry properties of the relevant band edges (located at the K, Q and $\Gamma$ points) of high-symmetry stackings with  0\degree (R-type) and 60\degree (H-type) twist angles. These R- and H-type stackings are the fundamental building blocks present in moiré and atomically reconstructed supercells, experimentally observed in samples with small twist angles. Our calculations reveal distinct hybridization signatures of the spin and orbital degrees of freedom as a function of the electric field and interlayer separation, depending on the energy band and on the stacking configuration. With this crucial knowledge of the spin-valley physics of the band edges, we expanded our analysis to the low-energy dipolar (interlayer) excitons, either direct or indirect in $k$-space. We found that these different dipolar exciton species present distinct electric dipole moments, polarizabilities, and valley Zeeman signatures (g-factors) due to the particular mixing of the spin and orbital angular momenta from their constituent energy bands. We also perform a symmetry analysis to discuss the phonon modes that could mediate the optical recombination of momentum-indirect dipolar excitons. Regarding the valley Zeeman signatures of the interlayer excitonic states, our calculations reveal that direct dipolar excitons at the K-valley carry a robust valley Zeeman effect nearly independent of the electric field. On the other hand, the momentum-indirect dipolar excitons in the H-type stackings, with valence band at the K valleys and conduction bands at the Q valleys, show a pronounced dependence of the valley Zeeman for positive values of applied electric fields. This peculiar g-factor dependence is unambiguously related to the spin-mixing of conduction bands at the Q point, which is absent in the R-type stackings. The valley Zeeman physics of the investigated dipolar excitons can also be affected by varying the interlayer distance, which could be experimentally probed under external pressure. As in the case of external electric fields, the interlayer variation effects in the valley Zeeman signatures are more pronounced in H-type stackings and are more likely to be observed in samples with twist angle close to 60$\degree$.

Although the \textit{intralayer} excitons were not in the scope of the present manuscript, these quasiparticles are still optically active and easily recognized in the photoluminescence or reflectivity spectra of \MS/\WS~heterobilayers\cite{Rivera2015NatComm, Nagler2017NatComm, Gobato2022NL}. Particularly, our g-factor calculations for the band edges (Fig.~\ref{fig:Efield_gz}) suggest that \textit{intralayer} excitons could also provide valuable information on the interlayer hybridization via the valley Zeeman as a function of the electric field. Gated devices based of \MS/\WS~heterostructures are within experimental reach\cite{Ciarrocchi2019NatPhot,Shanks2022NL,Ciarrocchi2022NatRevMat} and could be used to test our hypothesis. We note that, in order to properly evaluate the \textit{intralayer} exciton physics, it would be desirable to go beyond DFT calculations and employ the GW + Bethe-Salpeter formalism\cite{Deilmann2018NL, Deilmann2020PRL, Barre2022Science, Amit2022}, which is beyond the scope of this work. Nonetheless, since these \textit{intralayer} excitons are also quite localized in $k$-space\cite{Chernikov2014PRL, Wang2018RMP, FariaJunior2023}, the spin and orbital angular momenta of the Bloch bands should account for the majority of the hybridization effects.

We emphasize that the theoretical approach employed here in our study is quite robust and provides quantitative microscopic insights into the spin-valley physics of these van der Waals heterobilayers. This framework can certainly be extended to more complex multilayered van der Waals heterostructures that also host dipolar excitons, for instance, large-angle twisted MoSe$_2$/WS$_2$\cite{Mcdonnell2020, Volmer2022}, homobilayers\cite{Deilmann2018NL, Lin2021NatComm_bi}, trilayer hetero- and homo-structures\cite{Li2021, Feng2022, Zhang2022}, and MoSe$_2$/WS$_2$ heterobilayers\cite{Alexeev2019Nature, Tang2021NatNano, Gobato2022NL}.

%================================================================================

\acknowledgments

We thank Yaroslav Zhumagulov and Christian Schüller for helpful discussions and Sivan Refaely-Abramson for valuable insights into interlayer excitons. The authors acknowledge the financial support of the Deutsche Forschungsgemeinschaft (DFG, German Research Foundation) SFB 1277 (Project-ID 314695032, projects B07 and B11), SPP 2244 (Project No. 443416183) and of the European Union Horizon 2020 Research and Innovation Program under Contract No. 881603 (Graphene Flagship).

%================================================================================

\section*{Appendix A: Computational details}

The DFT are performed using the WIEN2k code\cite{wien2k}, which implements a full potential all-electron scheme with augmented plane wave plus local orbitals (APW+lo) method. The crystal structures of monolayers and the heterostructure are generated using the atomic simulation environment (ASE) python package\cite{ASE}. We used the Perdew-Burke-Ernzerhof exchange-correlation functional\cite{Perdew1996PRL}, a Monkhorst-Pack k-grid of 15$\times$15 and a self-consistent convergence criteria of 10$^{-6}$ e for the charge and 10$^{-6}$ Ry for the energy. For the heterostructures, van der Waals corrections3\cite{Grimme2010JCP} are taken into account in the self-consistent cycle. The core–valence energy separation is chosen as $-6$ Ry, the atomic spheres are expanded in orbital quantum numbers up to 10 and the plane-wave cutoff multiplied by the smallest atomic radii is set to 9. For the inclusion of SOC, core electrons are considered fully relativistically whereas valence electrons are treated in a second variational step\cite{Singh2006}, with the scalar-relativistic wave functions calculated in an energy window of -10 to 8 Ry. This upper energy limit provides a sufficiently large number of bands ($\sim$ 1500) to guarantee the proper convergence of the orbital angular momentum, as shown in Refs.~\cite{Wozniak2020PRB, Forste2020NatComm, FariaJunior2022NJP}. The applied out-of-plane electric field is included as a zig-zag potential added to the exchange-correlation functional\cite{Stahn2001PRB} with 40 Fourier coefficients. The structural parameters are discussed in Section~\ref{sec:gen}.

%================================================================================

\bibliography{biblio}

%===============================================================================

\end{document}